\documentclass[submission, Phys]{SciPost}

\pdfoutput=1

\usepackage[utf8]{inputenc} 
\usepackage[T1]{fontenc} 	
\usepackage[english]{babel} 

\usepackage[bitstream-charter]{mathdesign}
\usepackage{geometry} 		
\usepackage{amsmath} 		
\usepackage{amsthm} 		
\usepackage{mathtools} 		
\usepackage{float} 			
\usepackage{graphicx} 		
\usepackage{tabularx} 		
\usepackage{booktabs} 		
\usepackage{color, xcolor} 	
\usepackage{pdfpages} 		
\usepackage{extarrows} 		
\usepackage{multirow} 		
\usepackage{multicol} 		
\usepackage{caption} 		
\usepackage{subcaption} 	
\usepackage{enumitem} 		
\usepackage{setspace} 		
\usepackage{xspace} 		
\usepackage{ragged2e} 		
\usepackage{stackrel} 		
\usepackage{tikz} 			
\usetikzlibrary{shapes.geometric, angles, quotes}
\usepackage{braket} 		
\usepackage{bm} 			
\usepackage{tensor} 		
\usepackage{slashed} 		
\usepackage{siunitx} 		
\usepackage{lastpage} 		
\usepackage{cite} 			
\usepackage[normalem]{ulem} 
\usepackage{fontawesome} 	
\usepackage{tocloft} 		
\usepackage{titlesec} 		
\usepackage{doi} 			
\usepackage{hyperref} 		
\usepackage[most]{tcolorbox} 					
\usepackage[nameinlink, capitalize]{cleveref} 	
\usepackage[nottoc, notlot, notlof]{tocbibind} 	
\usepackage[ruled, vlined]{algorithm2e} 		
\usepackage{makecell}
\usepackage{wrapfig}

\hypersetup{
	pdftitle={Virtues and Vices of Equivariant Transformers},
	pdfauthor={Luigi Favaro, Tilman Plehn, Huilin Qu, and Jonas Spinner},
	colorlinks=true, 			
	linkcolor={red!50!black}, 	
	citecolor={blue!50!black}, 	
	urlcolor={blue!80!black} 	
} 

\DeclareSymbolFont{usualmathcal}{OMS}{cmsy}{m}{n}
\DeclareSymbolFontAlphabet{\mathcal}{usualmathcal}

\newlist{todolist}{itemize}{2}
\setlist[todolist]{label=$\square$}
\usepackage{pifont}
\SetArgSty{textnormal}
\SetKwComment{Comment}{{\small\#}~}{}
\SetCommentSty{mycommfont}

\setitemize{itemsep=0pt, parsep=0pt} 				
\setenumerate{itemsep=0pt, parsep=0pt} 				
\setitemize{itemsep=2pt,topsep=2pt,parsep=0pt,partopsep=0pt,leftmargin=*}
\setenumerate{itemsep=0pt,topsep=2pt,parsep=0pt,partopsep=0pt,labelindent=3pt,leftmargin=*}
\newcommand\one{\leavevmode\hbox{\small1\normalsize\kern-.33em1}}

\newcommand{\gev}{\text{GeV}}
\newcommand{\tev}{\text{TeV}}

\def\slashchar#1{\setbox0=\hbox{$#1$}           
   \dimen0=\wd0                                 
   \setbox1=\hbox{/} \dimen1=\wd1               
   \ifdim\dimen0>\dimen1                        
      \rlap{\hbox to \dimen0{\hfil/\hfil}}      
      #1                                        
   \else                                        
      \rlap{\hbox to \dimen1{\hfil$#1$\hfil}}   
      /                                         
   \fi}

\newcommand{\dz}{\phantom{0}}
\newcommand{\result}[2]{{#1}~\textcolor{error}{$\pm$ {#2}}}

\definecolor{error}{rgb}{0.7, 0.7, 0.7}

\usepackage{scalerel}

\graphicspath{{./figs/}}                

\begin{document}

\begin{flushright}
    \vspace*{-1cm}
    {\footnotesize IRMP-CP3-26-23}
    \vspace*{0.5cm}
\end{flushright}

\begin{center}{\Large \textbf{
Virtues and Vices of Equivariant Transformers
}}\end{center}

\begin{center}
Luigi Favaro\textsuperscript{1},
Tilman Plehn\textsuperscript{2,3},
Huilin Qu\textsuperscript{4,5},
and Jonas Spinner\textsuperscript{6}
\end{center}

\begin{center}
{\bf 1} Centre for Cosmology, Particle Physics and Phenomenology, CP3, UCLouvain, Belgium \\
{\bf 2} Institut für Theoretische Physik, Universität Heidelberg, Germany \\
{\bf 3} Interdisciplinary Center for Scientific Computing (IWR), Universit\"at Heidelberg, Germany \\
{\bf 4} State Key Laboratory of Dark Matter Physics, Tsung-Dao Lee Institute \& School of Physics and Astronomy, Shanghai Jiao Tong University, China \\
{\bf 5} Key Laboratory for Particle Astrophysics and Cosmology (MOE) 
\& Shanghai Key Laboratory for Particle Physics and Cosmology,
Shanghai Jiao Tong University, China\\
{\bf 6} Institute for Particle Physics Phenomenology, Durham University, United Kingdom
\end{center}

\begin{center}
\today
\end{center}

\section*{Abstract}
{
We study for the first time the benefit of Lorentz-equivariant transformers for large-size jet tagging and flavor tagging. To control their computing demands, we optimize all implementations for inference cost metrics. In our scaling studies, we find that Lorentz-equivariant networks outperform standard transformers, provided geometric features are relevant. This holds true in an idealized world as well as for limited resources. The conditional gain from Lorentz equivariance provides interesting input to the development of foundation models for LHC data.
}

\vspace{2pt}
\noindent\rule{\textwidth}{1pt}
\tableofcontents\thispagestyle{fancy}
\noindent\rule{\textwidth}{1pt}

\clearpage
\section{Introduction}
\label{sec:intro}

Modern machine learning (ML) is boosting all aspects of the LHC research program and will transform the research methodology in view of the High-Luminosity LHC (HL-LHC). As new methods require flagship applications, the most advanced and established ML applications to LHC physics are jet taggers~\cite{Kasieczka:2019dbj,Nachman:2022emq}. Their goal is to make use of all information about all constituents to determine the partonic nature and properties of a jet~\cite{Cogan:2014oua,Baldi:2014kfa,deOliveira:2015xxd,Gallicchio:2010sw,Kasieczka:2017nvn}. 

This real-life application leads directly to the kind of representation we choose for the input and for the latent representations. From initial convolutional networks on calorimeter images, language-inspired recursive networks~\cite{Louppe:2017ipp}, and theory-inspired representations~\cite{Komiske:2018cqr} we have observed a transition to powerful and permutation-invariant graph networks~\cite{Qu:2019gqs} and transformers~\cite{Qu:2022mxj}. Transformers are especially attractive because they benefit most from pre-training and the ever-increasing size of training datasets~\cite{Vigl:2026ppx}.

From a physics perspective it is clear that the performance of jet taggers will benefit from the incorporating known symmetries, most notably permutation invariance and the Lorentz symmetry. This motivates Lorentz-equivariant (or Lorentz-covariant) graph networks~\cite{Butter:2017cot,Gong:2022lye,Qiu:2022xvr,Bogatskiy:2022czk,Qiu:2023ihi,ruhe2023clifford}. The current state of the art in physics-enhanced network architectures is Lorentz-equivariant transformers with learned symmetry breaking, using a geometric algebra representation~\cite{Brehmer:2024yqw,Petitjean:2025zjf} or local reference frames per constituent~\cite{Favaro:2025pgz}. These equivariant transformers should be more stable to train and more expressive in their inference as they do not have to use parameters for general structures.

An alternative path to high-performance taggers is pre-training of large transformers on large datasets, where we assume that the learned latent representation reflects the physics symmetries~\cite{Qu:2022mxj,Brehmer:2024yqw,Favaro:2025pgz,Bhimji:2025isp,Birk:2025fbs}, as shown for regression transformers in Ref.~\cite{Villadamigo:2025our}. A challenge at the LHC is potential limitations in training dataset size, training compute, or evaluation memory. 

Because in LHC physics symmetries are typically broken by the experiment, this duality is softened in that transformers either assume a symmetry and learn their breaking or learn the symmetry from scratch. An interesting quantitative question is which of these two methods works better~\cite{Breso-Pla:2026tlz}. Related to these learned representations, the next step will be to understand how transformers construct their latent representations~\cite{Vent:2025ddm,Patel:2026zbq,Agarwal:2026uqw,Kuntz:2026kuv} and how this affects their performance and reliability in a physics setting~\cite{Flek:2026vof}.

The LHC experiments are developing transformers for tagging large-$R$ jets~\cite{CMS-DP-2026-104,ATL-PHYS-PUB-2026-013} and for flavor tagging~\cite{CMS-DP-2024-066,CMS-DP-2025-081,ATLAS:2025dkv,ATL-PHYS-PUB-2026-001,Barr:2025djz,ATLAS:2026vyw} separately. While large-$R$ jets and their mass drops from weak-scale decays are likely to benefit more from Lorentz-invariance, merging the two tasks and training large networks jointly might well boost the performance on both sides.  In this paper we unify advanced flavor and large-$R$ jet taggers for the first time. 

In Section~\ref{sec:tagging_networks} we present four different leading transformer architectures, three of them symmetry-aware, two equivariant with learned symmetry breaking. We present improved implementations, to put our comparison of performance vs cost on a solid footing. In Section~\ref{sec:all_datasets} we benchmark their performance on the ATLASTop dataset with realistic large-$R$ top jets~\cite{ATLAS:2024rua}, the JetClass dataset covering different electroweak decay jets~\cite{Qu:2022mxj}, and the ATLAS JetSet dataset for flavor tagging~\cite{ATLAS:2025dkv}. For all architectures and all datasets we study the performance scaling with the number of network parameters. Moving on to a more realistic setting, in Section~\ref{sec:fixed_costs} we compare the same architectures on the same datasets, but for limited energy consumption, inference time, and memory use. Finally, in Section~\ref{sec:pretraining} we pre-train the leading L-GATr-slim architecture on the JetClass dataset and compare its performance after fine-tuning on the classic top tagging dataset~\cite{Kasieczka:2019dbj}. We provide additional results in a series of appendices.

\section{LHC Transformers}
\label{sec:tagging_networks}

Motivated by their compelling performance especially when we include pre-training, we compare a set of transformers designed for LHC jet tagging,
\begin{itemize}
    \item Baseline transformer, similar to Salt/GN~\cite{Barr:2025djz,ATLAS:2025dkv,ATLAS:2026vyw};
    \item Particle Transformer (ParT)~\cite{Qu:2022mxj};
    \item Lorentz-equivariant Geometric Algebra Transformer (L-GATr)~\cite{Brehmer:2024yqw} and L-GATr-slim~\cite{Petitjean:2025zjf};
    \item Lorentz Local Canonicalization (LLoCa) Transformer~\cite{Favaro:2025pgz}.
\end{itemize}
All of them construct an appropriate latent representation for LHC physics, so they can be applied to object classification, amplitude regression, event generation, etc.~\cite{Brehmer:2024yqw,Favaro:2025pgz}. Our baseline transformer differs slightly from the Salt/GN architecture used by the ATLAS collaboration, as discussed in Appendix~\ref{app:gn3}.

\subsection{Architectures}

All four architectures are physics-inspired transformers. We focus on their physics aspect and unify physics-agnostic design choices, such as a common GELU activation function and gated linear units with an increase factor of two in the MLP. Comparing global token aggregation with mean aggregation, we find similar performance and adopt the global token aggregation everywhere. The class attention aggregation of ParT remains as a physics choice. 

At the LHC, the beam direction breaks the Lorentz symmetry to rotations around the beam axis with boosts along the beam axis, $\text{SO}(2)\times \text{SO}^+(1,1)$. The detector breaks the symmetry further, such that effectively no continuous symmetry is left after detector simulation. Correspondingly, all Lorentz-equivariant architectures have to allow for a learned symmetry breaking. Such a learned symmetry breaking reduces the fundamental difference between explicit and implicit equivariance to the more technical question if we want to learn the breaking of a known symmetry or learn the broken symmetry directly~\cite{Breso-Pla:2026tlz}. Our transformers incorporate Lorentz equivariance in three different ways:
\begin{itemize}
\item ParT uses pairwise features invariant under $\text{SO}(2)\times \text{SO}^+(1,1)$,
\begin{align}
    \left\{ m_{ij}^2, \quad
    \Delta R_{ij}, \quad 
    k_{T,ij}=\min (p_{T,i},p_{T,j})\Delta R_{ij}, \quad 
    z_{ij}=\frac{\min (p_{T,i},p_{T,j})}{p_{T,i}+p_{T,j}}
    \right\}  \; .
  \end{align}
They give the network the option of exact equivariance by ignoring the node features, in the spirit of a soft symmetry prior. In particular $m_{ij}^2$ improves training and increases expressivity~\cite{Butter:2017cot,Bogatskiy:2023nnw}.

Standard transformers do not include pairwise features because processing them scales quadratically rather than linearly with the particle number. Scaled dot-product attention is the only standard operation that scales quadratically, but it does not involve learnable operations and can be implemented efficiently via flash attention. ParT includes pairwise features to increase expressivity in an efficient manner. It adds an initial embedding using a smaller latent space to keep the cost of this operation similar to the linear layers on nodes. The embedded features are included as attention bias in the scaled dot-product attention at essentially zero cost. MIParT~\cite{Wu:2024thh} and IAFormer~\cite{Esmail:2025kii} re-compute them for every attention operation with separate embeddings for each attention head. This improves performance on small datasets at an increased memory cost. PETv2~\cite{Bhimji:2025isp} includes a local attention preprocessing. We use ParT as a compromise between performance and cost.

\item L-GATr operates on the irreducible representations~\cite{dehaan2023euclidean} of the Lorentz group, similar to the LorentzNet~\cite{Gong:2022lye} and CGENN~\cite{ruhe2023clifford} graph networks. Explicit reference vectors allow the transformer to break Lorentz equivariance down to no residual symmetries. L-GATr operates on the full 16-dimensional multivector representation of the geometric algebra and works with a geometric product that also includes an outer product between vectors. L-GATr-slim uses only a scalar product between vectors and the multiplication of a scalar with a vector. Both use the Minkowski inner product to compute the vector contribution to the attention logits, allowing them to trivially recover $m_{ij}^2$ using vectors that are already expressive learned embeddings.

\item LLoCa makes neural networks Lorentz-equivariant through local frames that turn features into Lorentz-invariants. The invariant features can be processed with any architecture, including the LLoCa-Transformer, LLoCa-ParT, or LLoCa-ParticleNet. The local frames are constructed by predicting a set of three vectors with a small Lorentz-equivariant FramesNet and then orthonormalizing them using the Gram--Schmidt algorithm in Minkowski space. A common frame for all particles is possible, but LLoCa achieves the highest performance for separate frames per particle. The attention between particles then includes a frame-to-frame transformation. We improve the original LLoCa implementation with a new variance-preserving rescaling in the attention mechanism, as discussed in Appendix~\ref{app:variance}.
\end{itemize}
Equivariant architectures typically boost performance at increased compute cost. A fair comparison requires a method to scale different architectures. We introduce a continuous parameter $s$ which controls the number of channels, blocks, attention heads, and the additional properties of equivariant networks. We show the parametrization and example integer values of $s$ in Table~\ref{tab:arch-scaling}. This allows us to compare jet taggers as extensions of a baseline transformer over three orders of magnitude in network parameter count. The symmetry-inspired ParT, L-GATr/L-GATr-slim, and LLoCa-Transformer only add parameters at the percent level, ParT by preprocessing pairwise features, L-GATr and L-GATr-slim through (multi-)vector channels, and LLoCa through the FramesNet.

\begin{table}[t]
\centering
    \begin{small} \begin{tabular}{l rrrrrr}
    \toprule
    & XS & S & M & L & XL & \\
    & $s=-2$ & $s=-1$ & $s=0$ & $s=1$ & $s=2$ & $s$ \\
    \midrule
    Channels & 32 & 64 & 128 & 256 & 512 & $2^{7+s}$ \\
    Blocks & 4 & 6 & 8 & 12 & 17 & $3\cdot 2^{\lfloor (s+3)/2\rfloor}$ \\
    Heads & 4 & 4 & 8 & 8 & 16 & $2^{3 + \lfloor s/2\rfloor}$\\
    \midrule
    L-GATr-slim vectors & 8 & 16 & 32 & 64 & 128 & $2^{5+s}$ \\
    L-GATr multivectors & 4 & 8 & 15 & 32 & 64 & $2^{4+s}$ \\
    L-GATr scalars & 24 & 48 & 96 & 192 & 384 & $3\cdot 2^{5+s}$ \\
    LLoCa Frames-Net channels & 8 & 16 & 32 & 64 & 128 & $2^{5+s}$ \\
    ParT edge features & 16 & 32 & 64 & 128 & 256 & $2^{6+s}$ \\
    ParT class-att'n blocks & 1 & 1 & 2 & 2 & 2 & 2 \\     \midrule

    Learning rate & $10^{-3}$ & $5\cdot 10^{-4}$ & $2.5\cdot 10^{-4}$ & $1.25\cdot 10^{-4}$ & $6.25\cdot 10^{-5}$ & $2.5\cdot 10^{-4}\cdot 2^{-s}$ \\
    Number of parameters ($\pm 5\%$) & $4.2\cdot 10^4$ & $2.5\cdot 10^5$ & $1.4\cdot 10^6$ & $7.9\cdot 10^6$ & $4.2\cdot 10^7$ & $1.4\cdot 10^6\cdot 2^{3/2\cdot s}$ \\
    FLOPs ($\pm 5\%$) & $6\cdot 10^6$ & $3\cdot 10^7$ & $2\cdot 10^8$ & $8\cdot 10^8$ & $4\cdot 10^9$ & $1.4\cdot 10^8\cdot 2^{3/2\cdot s}$ \\
    \bottomrule
    \end{tabular} \end{small}
    \caption{Scaling prescription for all jet tagging transformers.}
    \label{tab:arch-scaling}
\end{table}

The same learning rate for all architectures, scaled with the number of channels, achieves strong performance. We use a common weight decay value of $10^{-2}$ for all architectures and sizes. We validate these settings and find that they lie on an approximately optimal plateau that is similar for all tasks investigated in Section~\ref{sec:all_datasets}.

\subsection{Compute cost}

\begin{figure}[t]
    \includegraphics[width=0.325\linewidth,page=1]{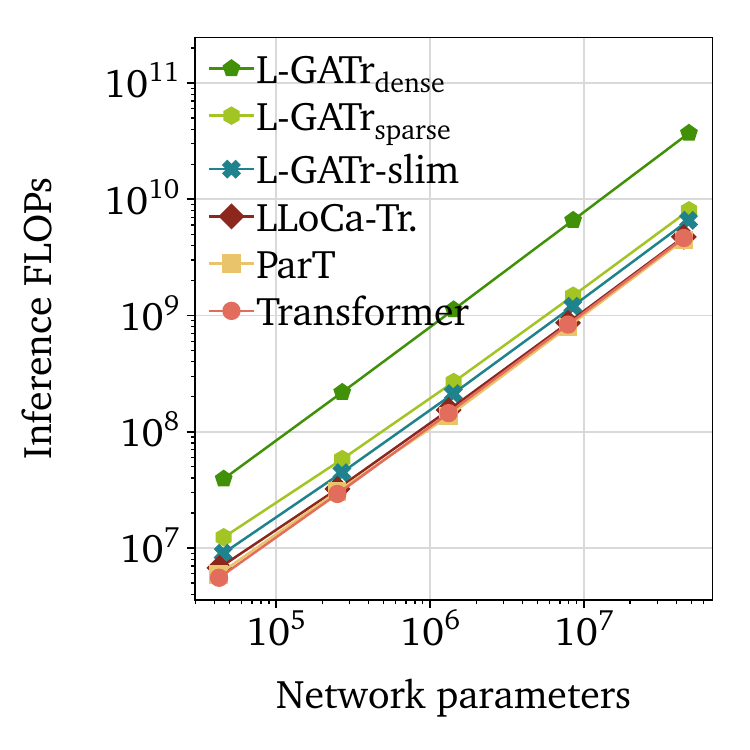}
    \includegraphics[width=0.325\linewidth,page=3]{figs/cost.pdf}
    \includegraphics[width=0.325\linewidth,page=5]{figs/cost.pdf} \\
    \includegraphics[width=0.325\linewidth,page=7]{figs/cost.pdf} 
    \includegraphics[width=0.325\linewidth,page=4]{figs/cost.pdf}
    \includegraphics[width=0.325\linewidth,page=6]{figs/cost.pdf}
    \caption{Inference cost metrics: FLOPs, CPU inference time and memory, GPU inference time and memory, and GPU energy use s a function of the network size.}
    \label{fig:cost}
\end{figure}

To quantify the computing cost, we consider a range of metrics evaluated on different devices:
\begin{enumerate}
    \item \textbf{Inference FLOPs \quad} Number of floating-point operations required to compute the tagging output of a jet with $N=50$ constituents. It is device-independent and can be calculated from the complexity of the layers. Training FLOPs are three times larger due to the backward pass.
    \item \textbf{CPU inference time \quad} Time for single-event processing with 50 constituents, the current default mode in ATLAS/CMS. Our numbers are based on a single Intel Xeon Platinum 8468V CPU, and use the Python \texttt{time.perf\_counter\_ns} tool, averaging over 100 steps.
    \item \textbf{CPU memory \quad} Running the inference pass on a separate CPU thread and use the \texttt{psutil} tool to extract memory use, again averaging over 100 steps.
    \item \textbf{Energy consumption \quad} Energy use to evaluate 50 constituents~\cite{Petitjean:2025zjf} on a NVIDIA A100 GPU in Joule, based on the energy cost estimates for elementary addition and multiplication operations in Refs.~\cite{6757323,wang2023bitnet}. This estimate assumes optimal GPU utilization, in practice the energy use can be $100\times$ larger.
    \item \textbf{GPU inference time \quad} For GPUs we consider batched event processing with batchsize 512, sampling events from the top tagging reference dataset. We use a single NVIDIA H100 GPU and \texttt{torch.cuda.Event}, averaging over 100 steps.
    \item \textbf{GPU memory \quad} We use \texttt{torch.cuda.max\_memory\_allocated} to measure the GPU memory use with batchsize 512, averaged over 100 steps. At this batchsize, only a small fraction of the 91GB GPU memory is used even for the largest networks.
\end{enumerate}
In Figure~\ref{fig:cost} we visualize their scaling with $s$. We focus on the inference cost. Training the standard taggers on the currently available public datasets takes at most a few days, whereas the inference requirements in experimental collaborations are highly constraining. Once larger datasets become available, the focus might shift to training time and memory use, as well as data efficiency. We show more results for training time and memory consumption in Appendix~\ref{app:cost}.

To motivate equivariant architectures based on their inference, we want to know whether the expressivity gain is worth the typically increased computational cost. Our scaling prescription allows us to compare architectures at fixed parameter count and also at fixed performance or fixed computational cost.

The initial implementation of a novel network architecture is typically not optimized for real-world comparisons. In contrast, widely used standard architectures come with highly efficient implementations. We implement established efficiency improvements for all architectures and  discuss three such directions and how they help all our architectures.

\subsubsection*{Compilation}

\begin{figure}[t]
    \includegraphics[width=0.325\linewidth,page=5]{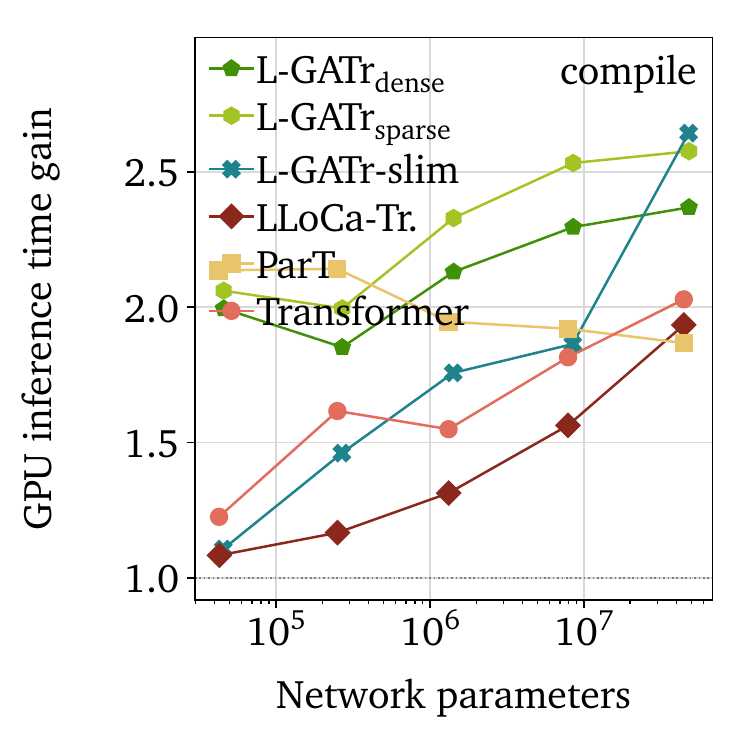}
    \includegraphics[width=0.325\linewidth,page=6]{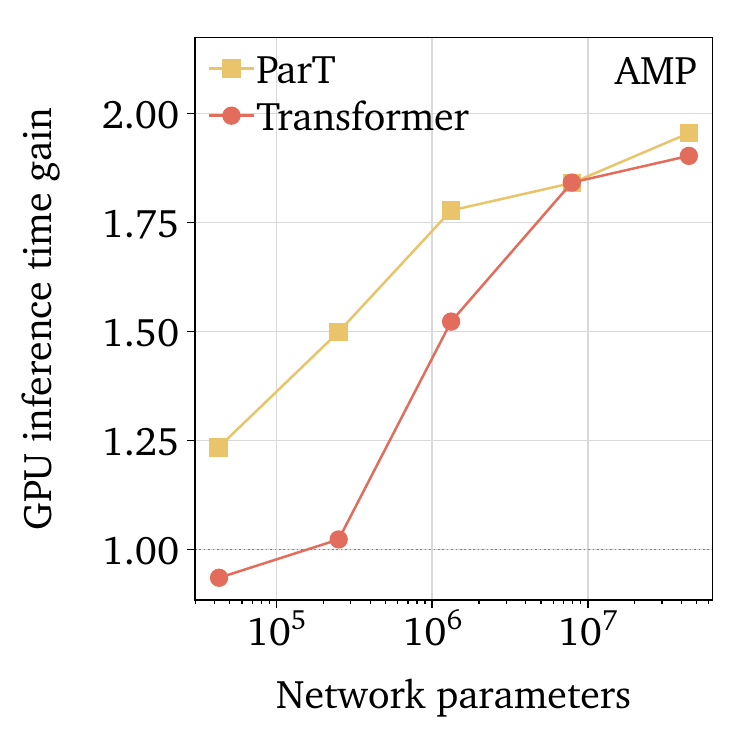}
    \includegraphics[width=0.325\linewidth,page=7]{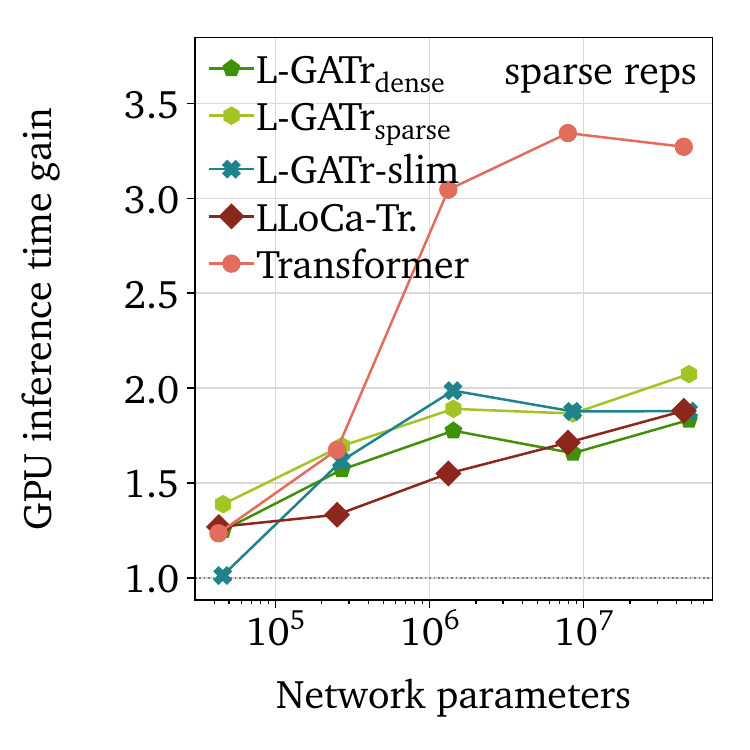}
    \caption{GPU inference time improvement from \texttt{torch.compile}, automatic mixed precision, and sparse jet representations. We show the speedup of the best combination for each of the three improvements.
      See Appendix~\ref{app:cost} for similar results on GPU memory and CPU time consumption.}
    \label{fig:cost_gains}
\end{figure}

PyTorch allows to compile a stack of operations into optimized code for CPU and GPU using \texttt{torch.compile}. It provides, for example, a selection of ideal kernels, kernel fusion, and joint optimization of forward and backward passes. In the left panel of Figure~\ref{fig:cost_gains}, we observe significant speedups for all architectures.

L-GATr and L-GATr-slim highly benefit from the compilation because they introduce new operations that are non-trivial to implement efficiently. The critical operations on multivector grades in L-GATr are the linear and geometric product operations. The original L-GATr implements them as dense matrix multiplications and uses zero-multiplications for forbidden operations. This dense implementation increases the number of operations, but is faster than a sparse implementation on GPUs because of the efficiency of the GEMM matrix multiplication kernels. With our new L-GATr implementation, the sparse variant is faster on geometric product operations. For linear operations, the dense approach is still faster on a GPU but uses fewer FLOPs and runs faster on a CPU. For this reason, we work with two versions based on their implementation of the linear layer
\begin{align}
  \text{L-GATr}_\text{dense}
  \qquad \text{vs} \qquad
  \text{L-GATr}_\text{sparse} \; . 
\end{align}
We find that $\text{L-GATr}_\text{dense}$ is faster on GPU, whereas $\text{L-GATr}_\text{sparse}$ is faster on CPU. Custom CUDA or triton kernels promise even more optimization. The potential gain is large for networks with sophisticated operations, such as equivariant networks, as recently demonstrated for $\text{E}(3)$-equivariant networks~\cite{cuequivariance,openequivariance}. 

\subsubsection*{Mixed precision}

By default, neural networks run in single (float32) or even double (float64) precision, but evaluation time and memory can be improved by using fewer bits. We only have to be careful not to lose accuracy in the network output.

Automatic mixed precision (AMP) allows for training most networks at float16 or bfloat16 precision. For inference-constrained applications, the current trend is towards int8 or int4 precision, or even 1-bit weights~\cite{wang2023bitnet}. For jet taggers, low-precision data types can be used at the cost of a small performance drop~\cite{Krause:2025qnl,Rai:2025cog}, also for Lorentz-equivariant networks~\cite{Petitjean:2025zjf}.

We use float32 as the default precision and aim for bfloat16 using AMP to reduce time and memory use. In Figure~\ref{fig:cost_gains} we see that the baseline transformer and ParT can be used with mixed precision without performance drop, although for the transformer at $s=-2$ AMP leads to a slight slowdown due to the data type conversions.

For the Lorentz-equivariant LLoCa-Transformer, L-GATr-slim, and L-GATr, training with AMP reduces performance at a rate that renders it impractical, so we stick to float32. This is different from our findings of Ref.~\cite{Petitjean:2025zjf}, which finds a small performance drop for the same architectures with bfloat16, but for a significantly smaller training dataset.

\subsubsection*{Sparse representation}

Jets are variable-length objects, exceeding 100 particles or particle-flow objects. To provide data in batches with variable particle number, the straightforward approach is zero padding. In transformers, these padded entries are masked out in attention, but they are still included in the latent representation, thus implying larger timing and memory costs. This wastes resources when more than one jet is processed at a time, such as during training or GPU inference.

A more efficient approach on the GPU is to concatenate all particles of the jets in a given batch size and process them in parallel, while keeping a pointer to the beginning of each jet. Avoiding zero padding reduces the inference time and memory use by around a factor two, as seen in Figure~\ref{fig:cost_gains}. The precise gain depends on the fraction of entries that would be zero-padded $r=N_\text{mean}/N_\text{max}$, with $N_\text{max}$ and $N_\text{mean}$ the maximum and average number of particles in the batch. They depend on the dataset distribution and chosen batch size. These variable-length input representations require specific attention kernels that implement the block-diagonal attention mask required to prevent information flow between different jets within the same batch. For PyTorch, these kernels are available, for instance, in the \texttt{flash-attn} and \texttt{xformers} library, and since recently also in native PyTorch. What is still needed is an efficient variable-length attention kernel implementation that supports the learnable attention bias used in ParT. We therefore use zero-padding for ParT and sparse jet representations for all other architectures.
\bigskip

\begin{table}[b!]
\centering
\begin{small} \begin{tabular}{lll rrr}
\toprule
Architecture & Accuracy & AUC & \text{Time} & \text{FLOPs} & \text{Memory} \\
\midrule
Baseline transformer \cite{Favaro:2025pgz} & 0.855 & 0.9867 & 15h $\to$ \dz{}9h & 210M & 2.3G \\
ParT \cite{Qu:2022mxj} & 0.861\dz{} & 0.9878\dz{} & 33h $\to$ 19h & 211M & 13.3G $\to$ \dz{}7.2G \\
$\text{L-GATr}_\text{sparse}$ \cite{Brehmer:2024yqw} & 0.865\dz{} & 0.9885\dz{} & 166h $\to$ 63h & 2060M $\to$ \dz{}352M & 19.0G $\to$ 16.8G \\
$\text{L-GATr}_\text{dense}$ \cite{Brehmer:2024yqw} & 0.865\dz{} & 0.9885\dz{} & 166h $\to$ 57h & 2060M $\to$ 1999M & 19.0G $\to$ 14.3G \\
L-GATr-slim \cite{Petitjean:2025zjf} & 0.866 & 0.9885 & 27h $\to$ 16h & 329M & 8.1G \\
LLoCa transformer \cite{Favaro:2025pgz} & 0.864 & 0.9882 & 28h $\to$ 12h & 219M & 4.1G\\
\bottomrule
\end{tabular} \end{small}
\caption{Cost metrics for the JetClass dataset, using the evaluation protocol of Refs.~\cite{Favaro:2025pgz,Petitjean:2025zjf}. Improvements are indicated by $\to$ and arise from \texttt{torch.compile}, the sparse geometric product in L-GATr, and micro-optimizations. All architectures run without AMP here, and all but ParT use sparse jet representations.}
\label{tab:jetclass_cost}
\end{table}

To summarize the efficiency gains from the improved implementation, we compare the new implementations with the results from Refs.~\cite{Favaro:2025pgz,Petitjean:2025zjf} on the JetClass dataset~\cite{Qu:2022mxj} in Table~\ref{tab:jetclass_cost}. All implementations are evaluated on a H100 GPU, after training for 1M iterations with batch size 512. The training time of all networks is reduced by values between 40\% and 70\%. For L-GATr the amount of FLOPs is reduced by 60M for $\text{L-GATr}_\text{dense}$ because of the sparse geometric product, and by an additional 1650M FLOPs for $\text{L-GATr}_\text{sparse}$ because of the sparse linear operations. The results in Figure~\ref{fig:cost} already include these improvements.

\section{Large-R jets and flavor tagging}
\label{sec:all_datasets}

We use three established public jet datasets with around 100M training jets to study the performance of the different tagging transformers:
\begin{itemize}
    \item ATLASTop: boosted top and QCD jets with full ATLAS simulation~\cite{ATLAS:2024rua};
    \item JetClass: ten classes of boosted objects with fast detector simulation~\cite{Qu:2022mxj};
    \item JetSet: flavor tagging with full ATLAS simulation~\cite{ATLAS:2025dkv}.
\end{itemize}
All networks are trained for 1M iterations with batch size 512, using a cosine annealing schedule with a warmup for the first 10\% iterations, and with the AdamW optimizer. The architectures and hyperparameters are given in Table~\ref{tab:arch-scaling}. We train networks with the five sizes $s=-2,-1,0,1,2$ in Table~\ref{tab:arch-scaling}, with the exception that we do not train L-GATr networks with $s=2$ due to the long training time. Trainings are performed on 1 NVIDIA A100-40GB GPU for $s\leq0$, while we use 4 NVIDIA A100-40GB or 4 H100 GPUs for the larger sizes. All trainings finish in less than 48 hours with the exception of L-GATr $s=1$ which instead takes $\approx 80$ hours.

To characterize the scaling of the various evaluation metrics $T$, we fit the function
\begin{align}
    \hat{T}(N, T_\infty, B, \beta) = T_\infty + \frac{B}{N^\beta} \; ,
\label{eq:scaling}
\end{align}
where $N$ is the number of learnable parameters or any other metric, and $(T_\infty, B, \beta)$ are fit parameters. In contrast to similar studies~\cite{Vigl:2026ppx}, we do not vary the dataset size. We perform the fit with the Huber loss, and using the L-BFGS optimizer. To estimate the fit uncertainty, we bootstrap the procedure 100 times and report an uncertainty band corresponding to the 0.1 quantile of the fit results.

In the remainder of this section, we show the fit results for each tagger as a function of the number of network parameters as a proxy of the network capacity. In Section~\ref{sec:fixed_costs} we track multiple cost metrics to determine the optimal architecture for a limited budget.

\subsection{ATLAS top tagging}

Using the ATLAS top tagging dataset, we test the benefits of Lorentz equivariance in a realistic setting with the full ATLAS detector simulation. The dataset also contains smaller samples with systematic uncertainty variations, allowing us to compare the resilience of the different architectures.

The nominal sample, used during training, contains 100M jets, equally divided between light-quark and top jets. They are generated at leading-order with Pythia8~\cite{Sjostrand:2014zea} using the standard ATLAS simulation chain. The top jets are extracted from the hypothetical decay 
\begin{align}
 Z' \rightarrow t\bar{t}
 \qquad \text{assuming} \quad m_{Z'}=2\, \tev \; .
\end{align}
The events are reweighted to a uniform distribution in 
\begin{align}
p_{T,j} = 0.35~\ldots~5\,\tev \; .
\end{align}
The dataset imposes various truth-level requirements for both classes; a detailed discussion of the cuts is included in Ref.~\cite{ATLAS:2024rua}. Fifteen additional samples contain systematic variations of the modeling of signal and background and of the experimental systematics related to tracking and energy clustering. Due to the small size of these samples, we only include a limited resilience study in Appendix~\ref{app:resilience}. Unlike the JetClass and JetSet datasets, the ATLAS top tagging dataset does not provide features beyond the particle 4-momenta.

\begin{figure}[t]
    \includegraphics[width=0.325\linewidth,page=1]{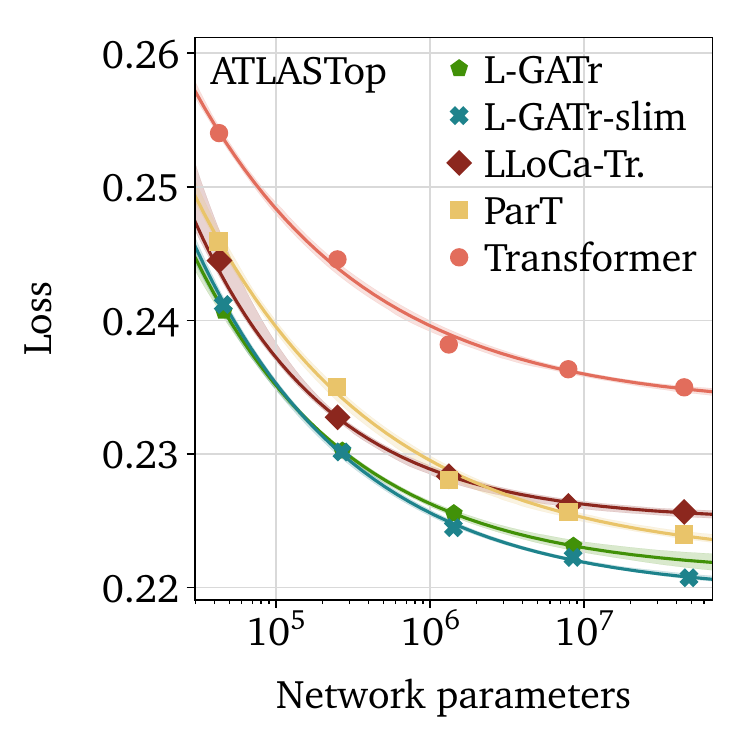}
    \includegraphics[width=0.325\linewidth,page=2]{figs/perf_atlastop5ep_final.pdf}
    \includegraphics[width=0.325\linewidth,page=4]{figs/perf_atlastop5ep_final.pdf}
    \caption{Performance on the nominal sample of the ATLAS top tagging dataset. We show scaling fits for the loss function value, AUC, and the background rejection at fixed signal efficiency.}
    \label{fig:perf_atlastop}
\end{figure}

In Figure~\ref{fig:perf_atlastop} we show the loss, the AUC, and the background rejection at fixed signal efficiency $0.5$ for the different architectures as a function of the network sizes. We do not observe any significant differences in hierarchy between the three evaluation metrics. First, symmetry-enhanced architectures clearly outperform the baseline transformer. Among them, the Lorentz-equivariant L-GATr and L-GATr-slim architectures achieve higher performance than the LLoCa-Transformer and the approximately equivariant ParT. L-GATr and L-GATr-slim behave almost identically, which indicates that the antisymmetric rank-2 tensor representation does not help for top tagging. The LLoCa-Transformer performance matches L-GATr and L-GATr-slim for small networks, but stalls towards large networks.

Altogether, these results confirm that the symmetry bias leads to significantly higher performance whenever geometric information dominates. In the absence of a simulation gap the network performance would still improve beyond 50M parameters.

\subsection{JetClass multi-class tagging}

\begin{figure}[t!]
    \includegraphics[width=0.325\linewidth,page=1]{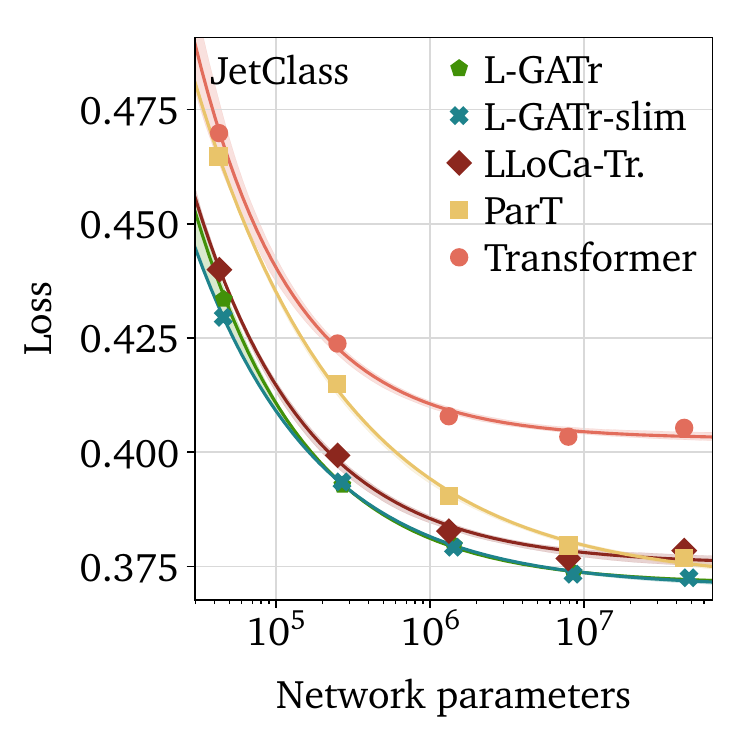}
    \includegraphics[width=0.325\linewidth,page=2]{figs/perf_jetclass5ep_final.pdf}
    \includegraphics[width=0.325\linewidth,page=3]{figs/perf_jetclass5ep_final.pdf} \\
    \includegraphics[width=0.325\linewidth,page=9]{figs/perf_jetclass5ep_final.pdf}
    \includegraphics[width=0.325\linewidth,page=11]{figs/perf_jetclass5ep_final.pdf} 
    \includegraphics[width=0.325\linewidth,page=12]{figs/perf_jetclass5ep_final.pdf} \\
    \includegraphics[width=0.325\linewidth,page=10]{figs/perf_jetclass5ep_final.pdf} 
    \includegraphics[width=0.325\linewidth,page=4]{figs/perf_jetclass5ep_final.pdf}
    \includegraphics[width=0.325\linewidth,page=5]{figs/perf_jetclass5ep_final.pdf} \\
    \includegraphics[width=0.325\linewidth,page=6]{figs/perf_jetclass5ep_final.pdf} 
    \includegraphics[width=0.325\linewidth,page=7]{figs/perf_jetclass5ep_final.pdf} 
    \includegraphics[width=0.325\linewidth,page=8]{figs/perf_jetclass5ep_final.pdf}
    \caption{JetClass performance scaling curves for the loss function, class-averaged AUC, accuracy, and background rejection rates at fixed signal efficiency for the 9 signal classes.
    }
    \label{fig:perf_jetclass}
\end{figure}

The JetClass training dataset includes 10 classes with 10M events each. The background class contains light-quark and gluon jets, while the signals are jets produced from electroweak $Z$, $W$, $H$, and top decays. The event generation follows the standard pipeline, MadGraph~\cite{Alwall:2014hca} for the hard scattering, Pythia8~\cite{Sjostrand:2014zea} for showering and hadronization, and Delphes3~\cite{deFavereau:2013fsa} for fast detector simulation. Jets are clustered using FastJet~\cite{Cacciari:2011ma} with the anti-$k_T$ algorithm with $R=0.8$. For the signal classes, fully reconstructed signatures are required. The kinematic cuts on reconstructed jets are
\begin{align}
p_{T,j} = 500~\ldots~1000\; \gev \qquad \text{and} \qquad |\eta_j| < 2.0 \; .
\end{align}
In addition to the constituent 4-momenta, the dataset provides the particle identifications (PIDs) and trajectory displacement variables. The former are scalar information from the point of view of Lorentz-equivariant taggers. The trajectory displacement variables contain geometric information, but the format is not suitable as a 4-vector in the Lorentz-equivariant architectures, so we also embed them as scalar features.

In the top row of Figure~\ref{fig:perf_jetclass} we show the scaling curves for the loss function value on the test dataset, the average AUC for the different binary hypothesis tests, and the accuracy. Again, we see a clear advantage of Lorentz-equivariant architectures, but the gap to the baseline transformer is smaller than for pure top tagging. This shows how all architectures benefit equally from the additional scalar features. For the symmetry-enhanced architectures, L-GATr and L-GATr-slim again lead, and the LLoCa-Transformer matches them for small networks. ParT catches up for large networks, ultimately matching the LLoCa-Transformer.

In the lower rows of Figure~\ref{fig:perf_jetclass} we show fits to the background rejection for all classes. We observe a similar hierarchy related to the degree of Lorentz equivariance, but the gain from Lorentz equivariance differs between the different classes. The wide performance gap for hadronic top, $Z$, and $W$ tagging is expected from the mass drop in the electroweak decay to quarks. Tagging leptonic top decays only involves a well-defined mass drop in the transverse mass, which is not available for tagging, so equivariant taggers struggle to benefit from symmetric information.  

For Higgs decays, Lorentz equivariance hardly helps with the double-$b$ tagging, since the distinctive $b$-jet signature, such as the presence of displaced tracks, already enables efficient tagging. For $H \to c\bar{c}$ decays, since charm tagging is much more challenging than $b$-tagging, Lorentz equivariance becomes more relevant for exploiting jet substructure information. The same is true for the light-jet Higgs decays, both 2-body and 4-body. 

\begin{figure}[t]
    \includegraphics[width=0.485\linewidth,page=1]{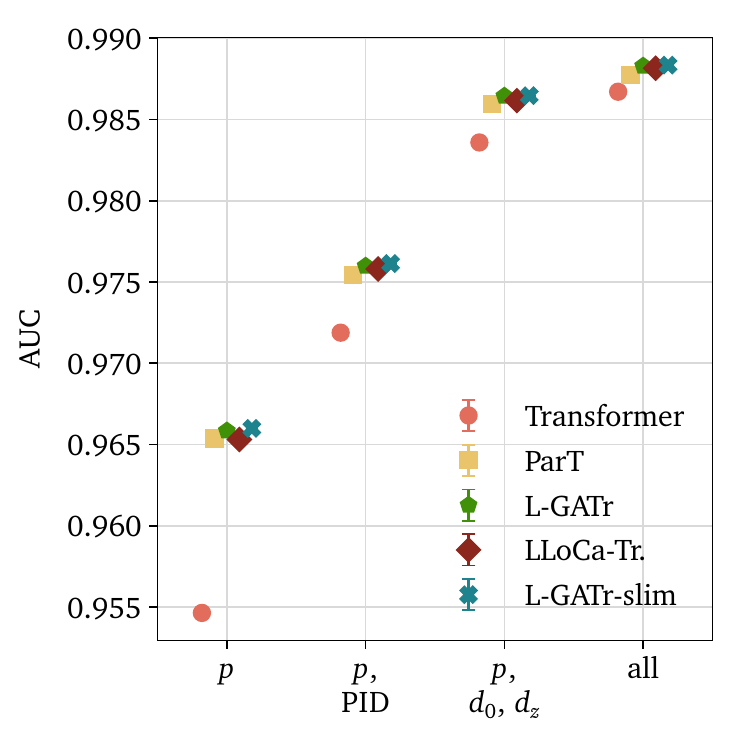}
    \includegraphics[width=0.485\linewidth,page=3]{figs/features.pdf}
    \caption{Relevance of input features on JetClass (left) and JetSet (right) for taggers of size $s=0$. We start with 4-vectors and add particle identification (PID) and trajectory displacement $(d_z)$ for JetClass. For JetSet, we add observables relevant to flavor tagging further described in the text.}
    \label{fig:feature_relevance}
\end{figure}

Next, we study the relevance of the additional particle type and trajectory displacement features in the left panel of Figure~\ref{fig:feature_relevance}. The gain from Lorentz-equivariant architectures is largest when only particle momenta are included. Particle identification or trajectory displacement helps Lorentz-equivariant and symmetry-agnostic architectures alike. With increasing performance, the required effort increases, so the differences between the architectures remain significant even when all features are included, see Figure~\ref{fig:perf_jetclass}. Even though the selection of the input features becomes more relevant than the encoded symmetries, the Lorentz-equivariant architectures achieve the best performance.

\subsection{JetSet flavor tagging}

\begin{figure}[t!]
    \includegraphics[width=0.325\linewidth,page=1]{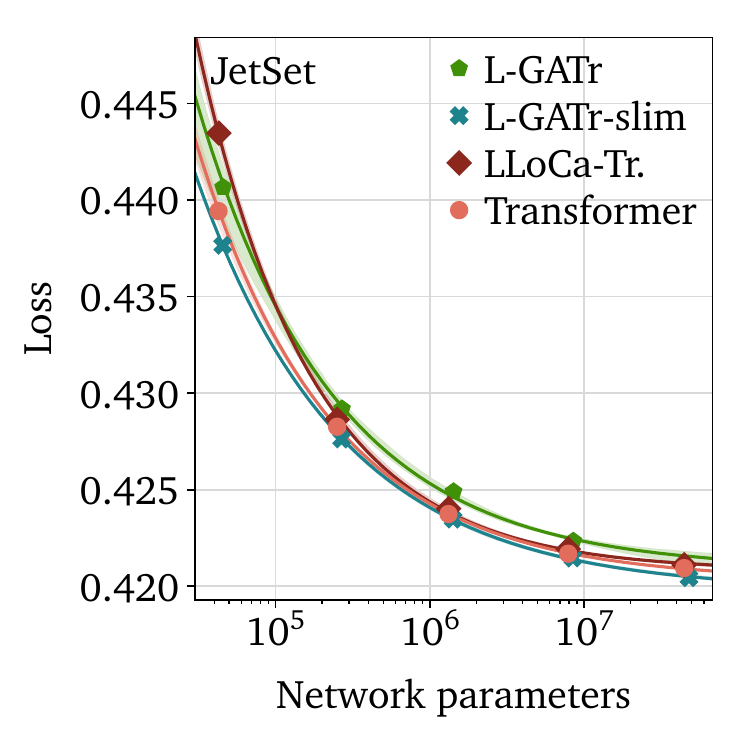}
    \includegraphics[width=0.325\linewidth,page=2]{figs/perf_jetset5ep_final.pdf}
    \includegraphics[width=0.325\linewidth,page=3]{figs/perf_jetset5ep_final.pdf} \\
    \includegraphics[width=0.325\linewidth,page=4]{figs/perf_jetset5ep_final.pdf}
    \includegraphics[width=0.325\linewidth,page=5]{figs/perf_jetset5ep_final.pdf}
    \includegraphics[width=0.325\linewidth,page=6]{figs/perf_jetset5ep_final.pdf} \\
    \includegraphics[width=0.325\linewidth,page=7]{figs/perf_jetset5ep_final.pdf}
    \includegraphics[width=0.325\linewidth,page=8]{figs/perf_jetset5ep_final.pdf}
    \includegraphics[width=0.325\linewidth,page=9]{figs/perf_jetset5ep_final.pdf}
    \caption{JetSet performance scaling curves for the loss function, class-averaged AUC, accuracy, and background rejection rates at fixed signal efficiency for the 6 signal/background pairs.}
    \label{fig:perf_jetset}
\end{figure}

For flavor tagging, we use the JetSet~\cite{ATLAS:2025dkv} dataset, part of the ATLAS
simulated data produced for the training of the GN2 tagger. The public dataset contains SM $t\bar{t}$ events generated to NLO with the Powheg Box v2~\cite{Alioli:2010xd} with NNPDF3.0~\cite{NNPDF:2014otw} parton distribution functions. Showering and hadronization is done by Pythia8 with the ATLAS tune A14. This dataset differs significantly from the other two large-R datasets. The kinematic range of the jets is limited to
\begin{align}
 p_{T,j}=20~...~3000\;\gev \; ,
\end{align}
and the maximum number of constituents is 40 rather than the 128 in the previous datasets. In addition to the reconstructed 4-momenta, we now have access to a wide range of additional features, all treated as scalars:
\begin{itemize} 
\item lifetime-signed track $d_0$ and the lifetime-signed longitudinal impact parameter $z_0\sin\theta$, with uncertainties, and encoded as significances $S_{d_0}=d_0/\sigma_{d_0}$ and $S_{z_0}=z_0/\sigma_{z_0\sin\theta}$; 
\item pure $d_0$ and $z_0\sin\theta$ variables;
\item the angular distance $\Delta R$ to the jet momentum;
\item ratio of the track charge and the momentum $q/|\vec p|$
\item constituent $\phi$ and $\eta$ with respective uncertainties;
\item number of hits in the detector trackers, which constitute nine additional variables.
\end{itemize}
In total, these are 19 scalar features used by ATLAS to train their taggers, which we preprocess in precisely the same way. However, we do not include the auxiliary tasks which we expect to boost the performance of all the architectures. 

We compare the performances in Figure~\ref{fig:perf_jetset}, again showing test loss, AUC and accuracy. We also show the $b$-tagging and $c$-tagging background rejections against the other jet classes. The outcome is very different from before, in that all architectures achieve similar performance. This is because the relevant tagging features are not the 4-momenta, but the track significance variables implemented as scalars. We exclude ParT from this comparison because the large number of scalar features is not foreseen in the current implementation. 

We show the impact of the different features in detail in the right panel of Figure~\ref{fig:feature_relevance}. We show the AUC after including the significance of the impact parameters (sig.), the impact parameters themselves (displ.), the additional track momentum variables (kin.), and the full set of scalars (all). We contrast their impact with the JetClass results and find that the performance now plateaus rapidly also in the 4-momenta only case. This drives the comparable performance of all architectures. 

\begin{table}[t]
\centering
\begin{small} \begin{tabular}{lcccccc}
  \toprule
  & \multicolumn{3}{c}{Exponent $\beta$} & \multicolumn{3}{c}{Asymptote $L_\infty$} \\
  & ATLASTop & JetClass & JetSet & ATLASTop & JetClass & JetSet \\
  \midrule
  Transformer & $0.39_{-0.03}^{+0.05}$ & $0.70_{-0.07}^{+0.06}$ & $0.50_{-0.02}^{+0.02}$ & $0.2335_{-0.0006}^{+0.0005}$ & $0.4030_{-0.0014}^{+0.0010}$ & $0.4203_{-0.0001}^{+0.0001}$ \\
  ParT & $0.38_{-0.04}^{+0.04}$ & $0.45_{-0.01}^{+0.02}$ & \textemdash & $0.2222_{-0.0007}^{+0.0007}$ & $0.3716_{-0.0009}^{+0.0010}$ & \textemdash \\
  LLoCa-Tr. & $0.52_{-0.05}^{+0.08}$ & $0.59_{-0.04}^{+0.06}$ & $0.59_{-0.03}^{+0.02}$ & $0.2251_{-0.0004}^{+0.0003}$ & $0.3754_{-0.0019}^{+0.0013}$ & $0.4208_{-0.0002}^{+0.0001}$ \\
  L-GATr & $0.43_{-0.04}^{+0.04}$ & $0.59_{-0.06}^{+0.03}$ & $0.47_{-0.07}^{+0.08}$ & $0.2210_{-0.0009}^{+0.0009}$ & $0.3710_{-0.0009}^{+0.0004}$ & $0.4207_{-0.0008}^{+0.0007}$ \\
  L-GATr-slim & $0.42_{-0.04}^{+0.03}$ & $0.55_{-0.02}^{+0.02}$ & $0.46_{-0.02}^{+0.01}$ & $0.2196_{-0.0004}^{+0.0004}$ & $0.3706_{-0.0008}^{+0.0004}$ & $0.4198_{-0.0002}^{+0.0002}$ \\
  \bottomrule
\end{tabular}
\end{small}
\caption{Exponent $\beta$ and asymptotic test loss $L_\infty$ of the scaling functions in Eq.\eqref{eq:scaling}. We report the median and 10\%/90\% quantile uncertainties from our bootstrapping approach.}
\label{tab:scaling_exponents}
\end{table}

Finally, we report the scaling exponent $\beta$ and the asymptotic loss value $L_\infty$ defined in Eq.~\eqref{eq:scaling} in Table~\ref{tab:scaling_exponents}, for all three datasets. The $L_\infty$ term indicates the best reachable loss with the available data. For the ATLASTop and JetClass datasets, there is a visible gap between the baseline transformer and the symmetry-aware architectures. The difference between ParT, LLoCa-Transformer, and L-GATr is within uncertainties, whereas L-GATr-slim extends to slightly better loss values. On the JetSet data, all networks converge to similar loss values, and only L-GATr-slim shows a marginal advantage. 

Given an $L_\infty$, the scaling exponent $\beta$ indicates how efficiently a network reaches its asymptotes. We see that ParT has the smallest convergence rate on the ATLASTop and JetClass datasets, and the other Lorentz-equivariant networks share similar scaling exponents. In particular, the baseline transformer saturates very quickly to a sub-optimal loss value on JetClass data, as indicated by the largest scaling exponent of $\beta = 0.70$. On the JetSet dataset, the LLoCa-Transformer shows the largest exponent, while the baseline transformer and L-GATr are within uncertainties. L-GATr-slim has an even smaller convergence rate which correlates with the smaller loss at infinity.

\clearpage
\section{Performance at fixed evaluation cost}
\label{sec:fixed_costs}

After discussing the raw performance, we now focus on the inference metrics introduced in Section~\ref{sec:tagging_networks}. We start with metrics related to the implementation, like FLOPs as a theory-based metric and the GPU energy use. As a large part of the LHC computing infrastructure will be stuck with CPUs, we benchmark the inference times on CPUs and GPUs and the memory consumption. We use the scaling fit procedure from Section~\ref{sec:all_datasets}, but evaluated on cost metrics instead of network size, and follow two approaches:
\begin{enumerate}
\item fix a cost-metric budget and determine which architecture performs best;
\item fix a target performance and see which architecture is the most cost-economic. 
\end{enumerate}
We simplify the discussion by choosing the test loss as a performance metric but observe similar behavior for AUC, accuracy, and background rejection. In each of the following figures, the best performing tagger defines a Pareto frontier, and we indicate the sub-optimal regions with a shaded background.

\subsubsection*{Energy consumption}

\begin{figure}[b!]
    \includegraphics[width=0.325\linewidth,page=2]{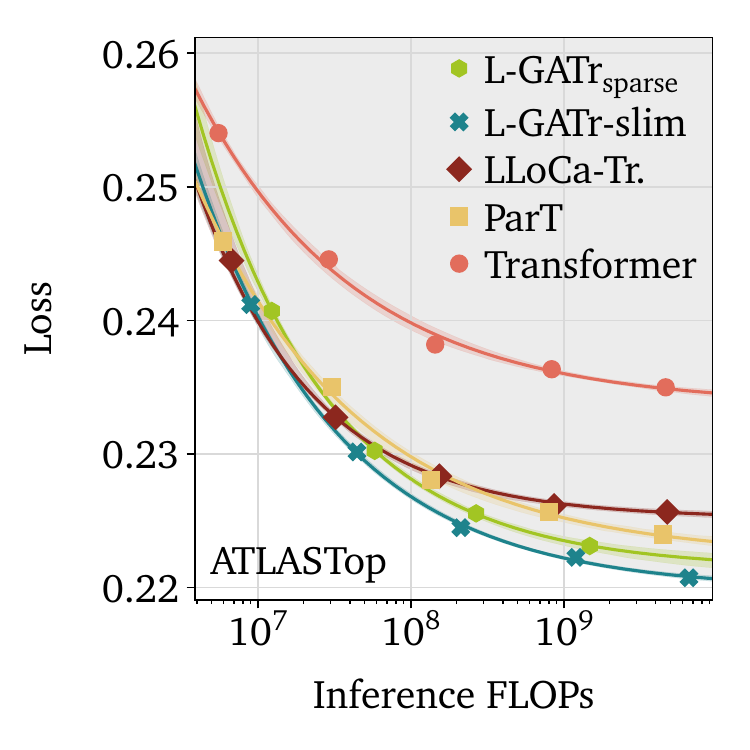}
    \includegraphics[width=0.325\linewidth,page=2]{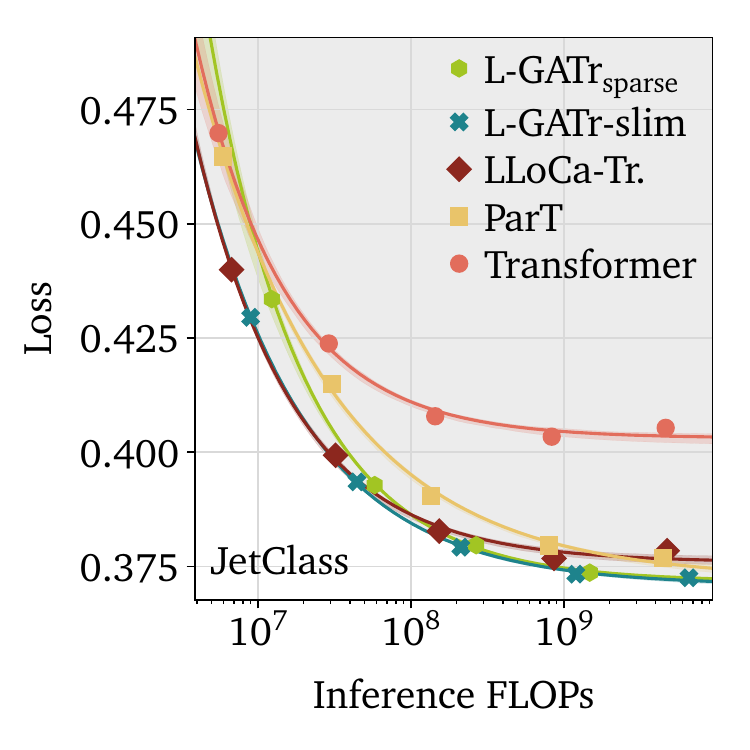}
    \includegraphics[width=0.325\linewidth,page=2]{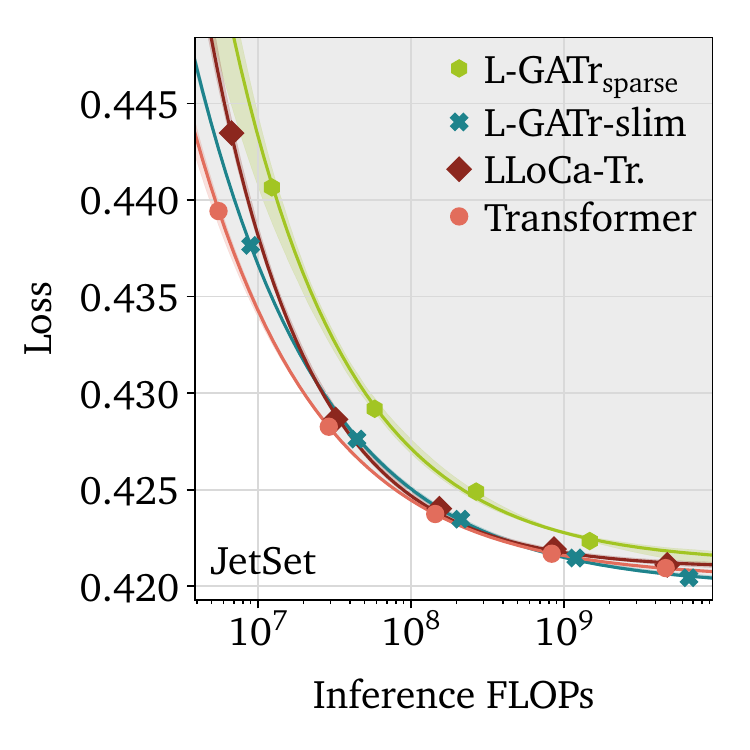} \\
    \includegraphics[width=0.325\linewidth,page=4]{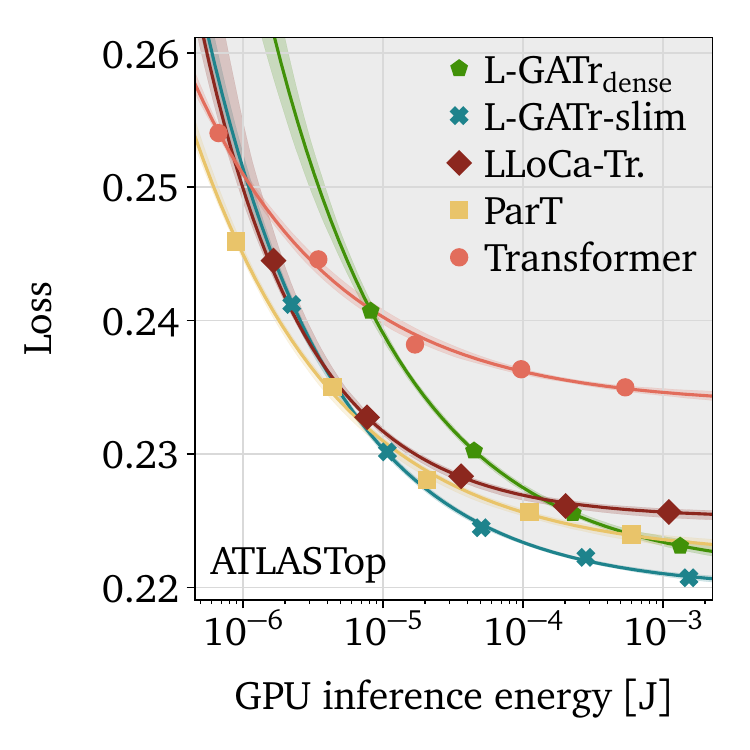}
    \includegraphics[width=0.325\linewidth,page=4]{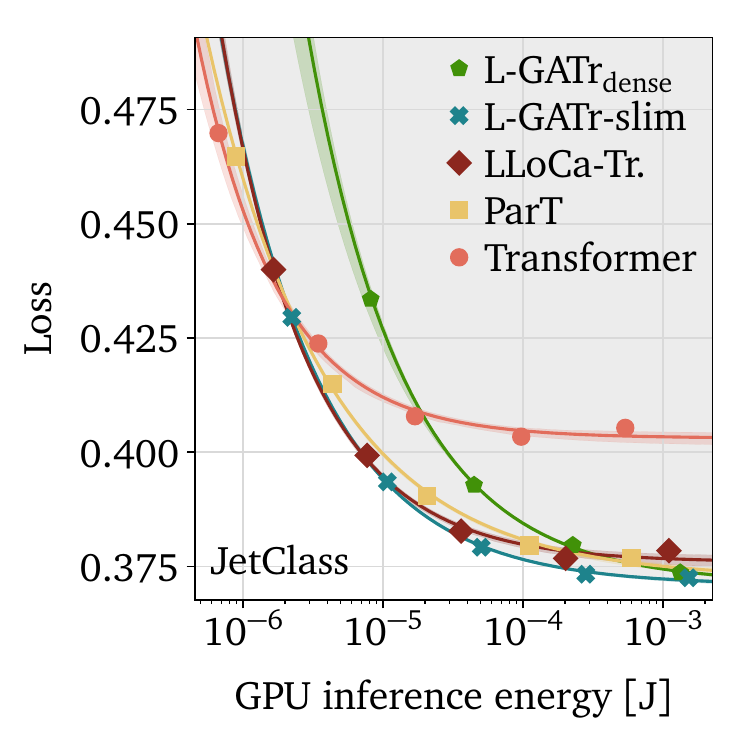}
    \includegraphics[width=0.325\linewidth,page=4]{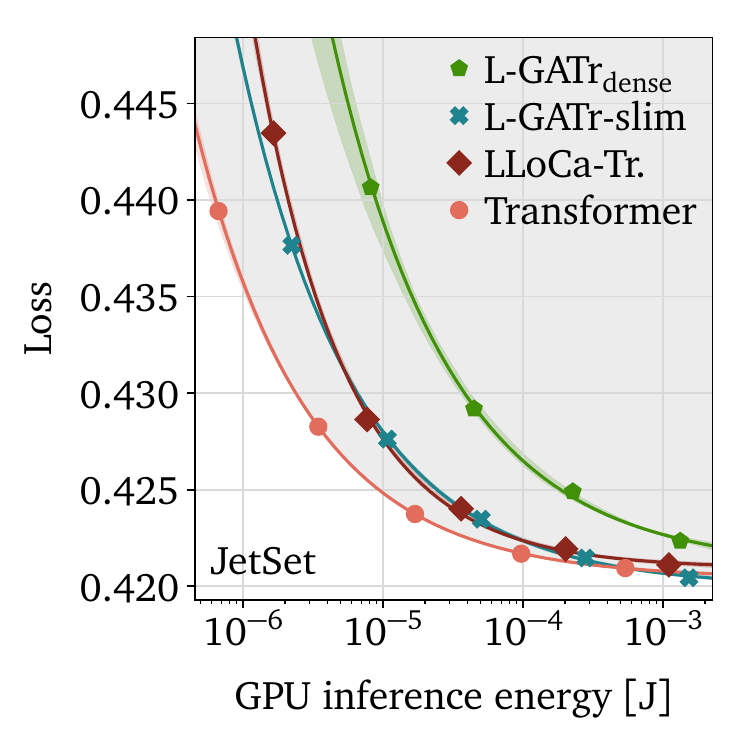}
    \caption{Inference FLOPs (top) and energy consumption on GPU (bottom) for the studied architectures. From left to right, we show the test loss over the inference metrics for ATLASTop, JetClass, and JetSet.}
    \label{fig:cost_1}
\end{figure}

First, we show the FLOPs and the energy cost on GPUs in Figure~\ref{fig:cost_1}, each column corresponding to a dataset. On ATLASTop and JetClass data L-GATr-slim leads on the Pareto front for both metrics. At low cost, all symmetry-aware architectures behave similarly and largely outperform the baseline transformer. The $\text{L-GATr}_\text{dense}$ energy use is larger because of the zero-multiplications. The scaling to larger networks is similar for all the architectures, giving an edge to L-GATr-slim.

On the JetSet dataset, the baseline transformer matches the performance of the other architectures. Due to the small overheads of the other architectures, it resides on the Pareto front in terms of energy consumption. However, these overheads require minimal additional FLOPs and are irrelevant for large networks, making L-GATr-slim and LLoCa-Tranformer viable alternatives.

\subsubsection*{Inference time}

\begin{figure}[t]
    \includegraphics[width=0.325\linewidth,page=5]{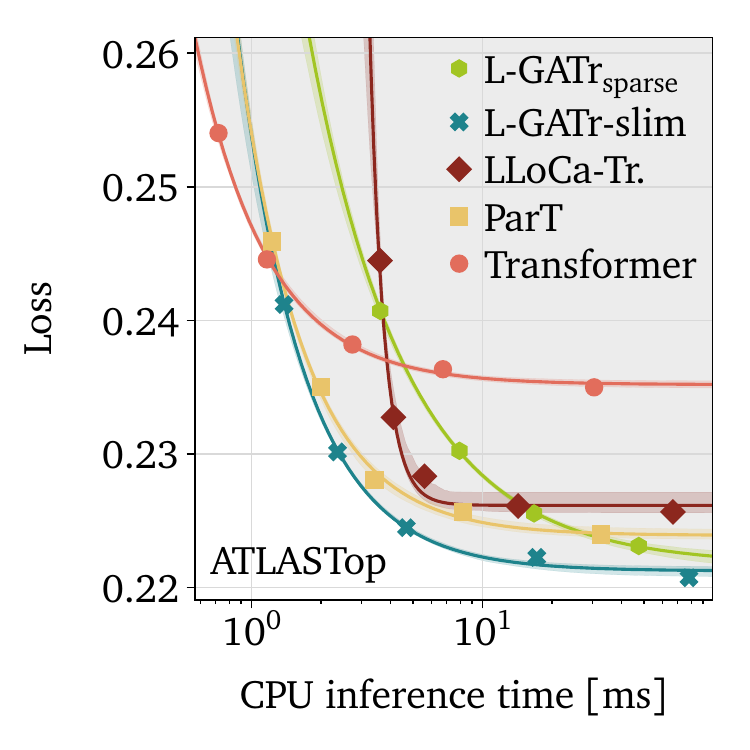}
    \includegraphics[width=0.325\linewidth,page=5]{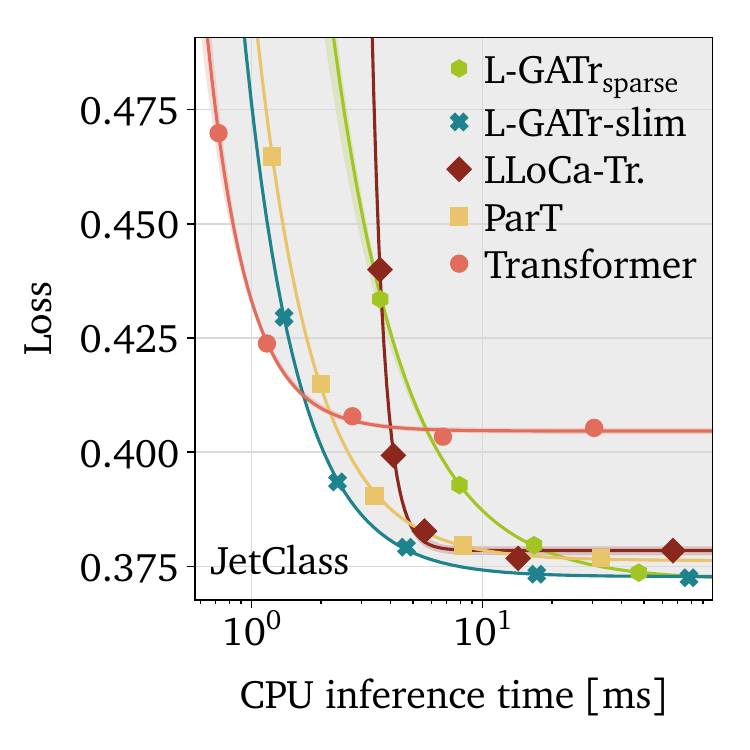}
    \includegraphics[width=0.325\linewidth,page=5]{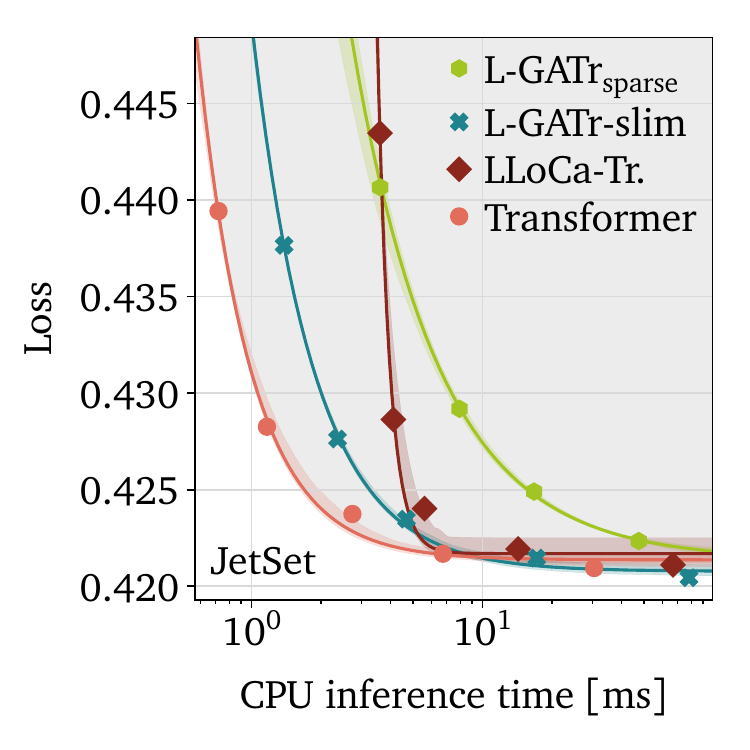} \\
    \includegraphics[width=0.325\linewidth,page=7]{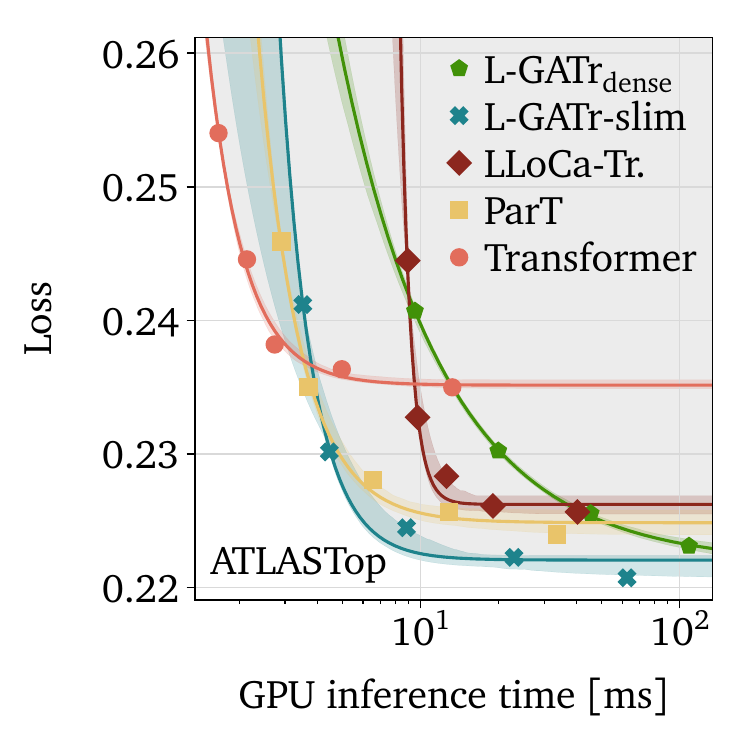}
    \includegraphics[width=0.325\linewidth,page=7]{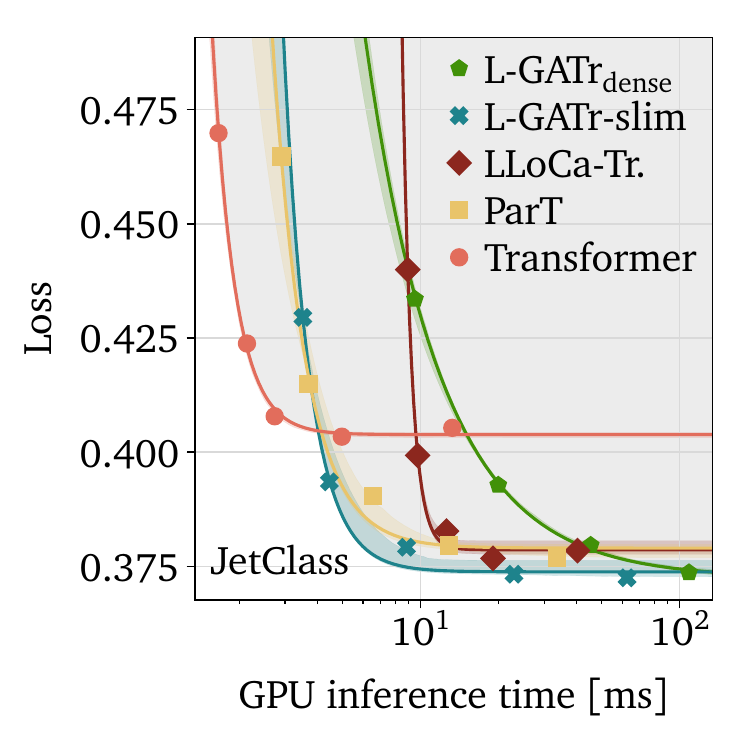}
    \includegraphics[width=0.325\linewidth,page=7]{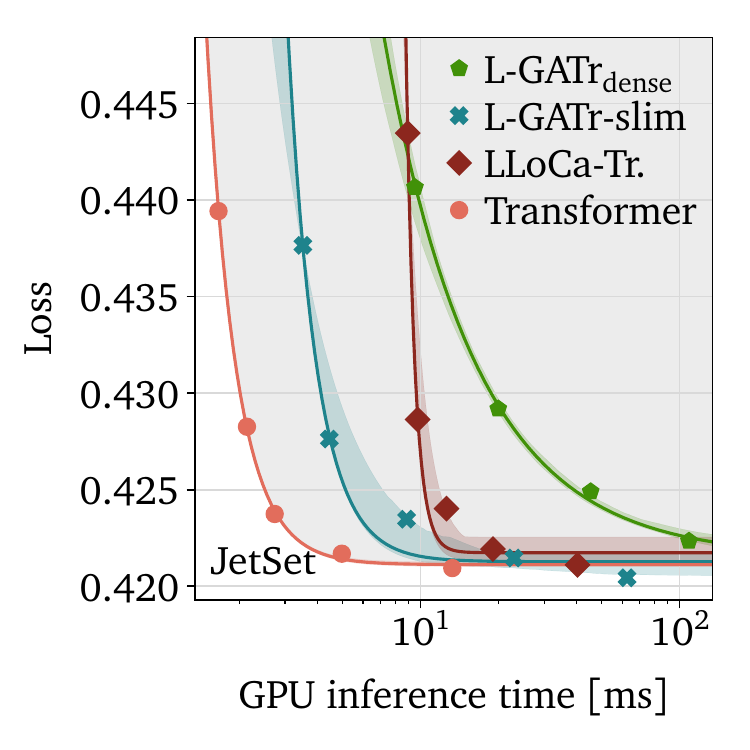}
    \caption{CPU (top) and GPU (bottom) inference time for the studied architectures. From left to right, we show the test loss over the inference metrics for ATLASTop, JetClass, and JetSet.}
    \label{fig:cost_2}
\end{figure}

Second, we show CPU and GPU inference times in Figure~\ref{fig:cost_2}. On our CPU, the baseline transformer provides the best performance for inference time $\leq 1 \,\text{ms}$ on the ATLASTop and JetClass datasets. For longer times, L-GATr-slim takes over, providing the best tradeoff between performance and inference speed.

The overhead for a small LLoCa-Transformer is caused by the frame prediction, which is extremely cheap in terms of FLOPs but uses less optimized code than the standard network operations. This becomes irrelevant after the intersection with the baseline transformer around $5\,\text{ms}$ on CPU and $10\,\text{ms}$ on GPU. The general L-GATr lags behind, as its operations are more difficult to implement efficiently.

For a sufficiently large compute budget, all equivariant architectures beat the baseline transformer. Our observations on JetSet are similar to the previous cost metrics: the baseline transformer provides the fastest inference and looses to the other architectures when the time budget exceeds $10\,\text{ms}$.

\subsubsection*{Memory}

Last, we discuss the memory use of the various architectures. As for the inference time, a memory limitation can be caused by a practical constraint on the available resources or the underlying framework. 

\begin{figure}[t]
    \includegraphics[width=0.325\linewidth,page=6]{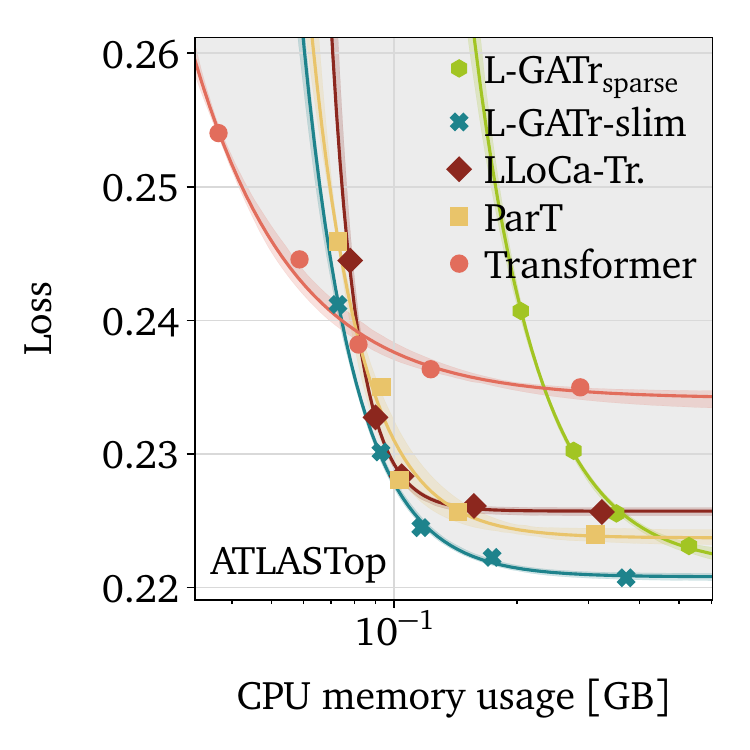} 
    \includegraphics[width=0.325\linewidth,page=6]{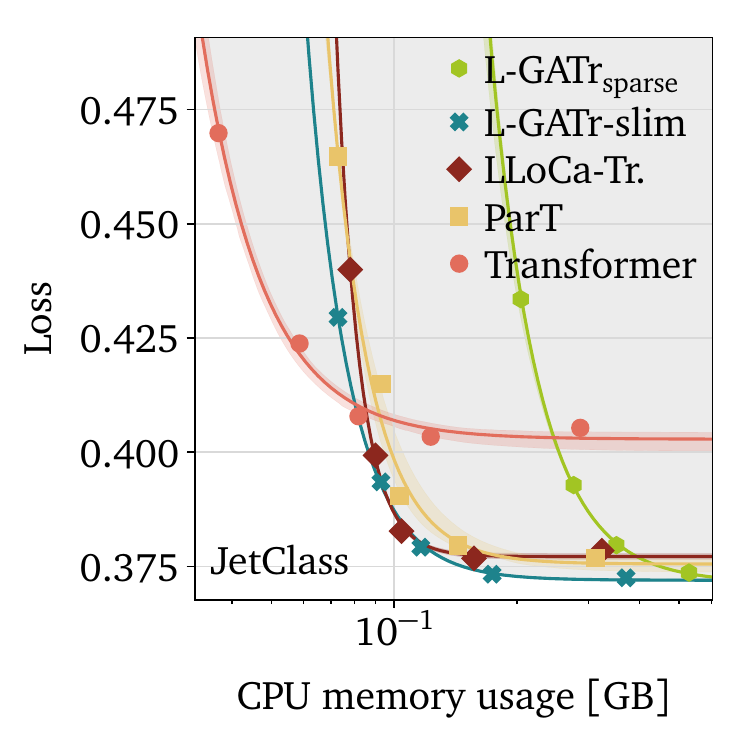}
    \includegraphics[width=0.325\linewidth,page=6]{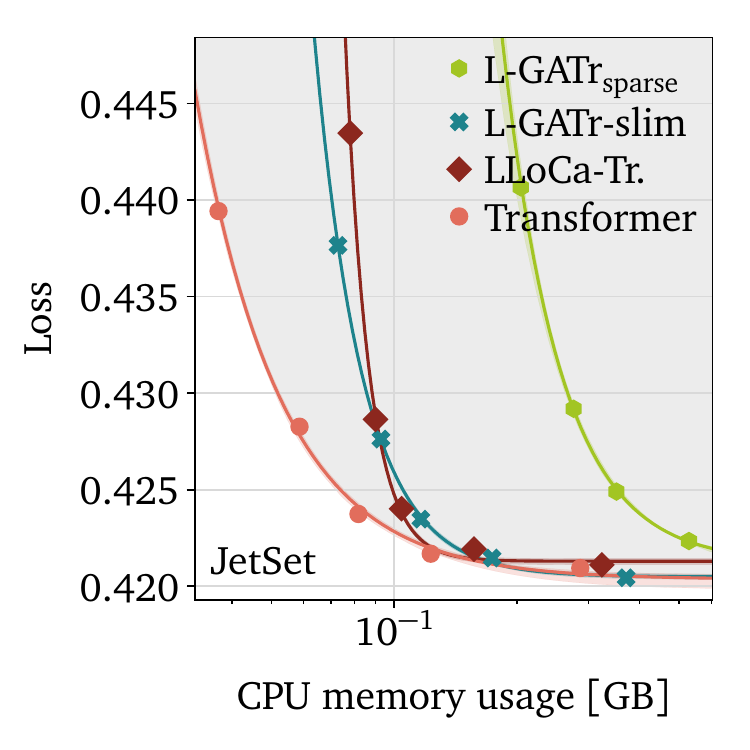} \\
    \includegraphics[width=0.325\linewidth,page=8]{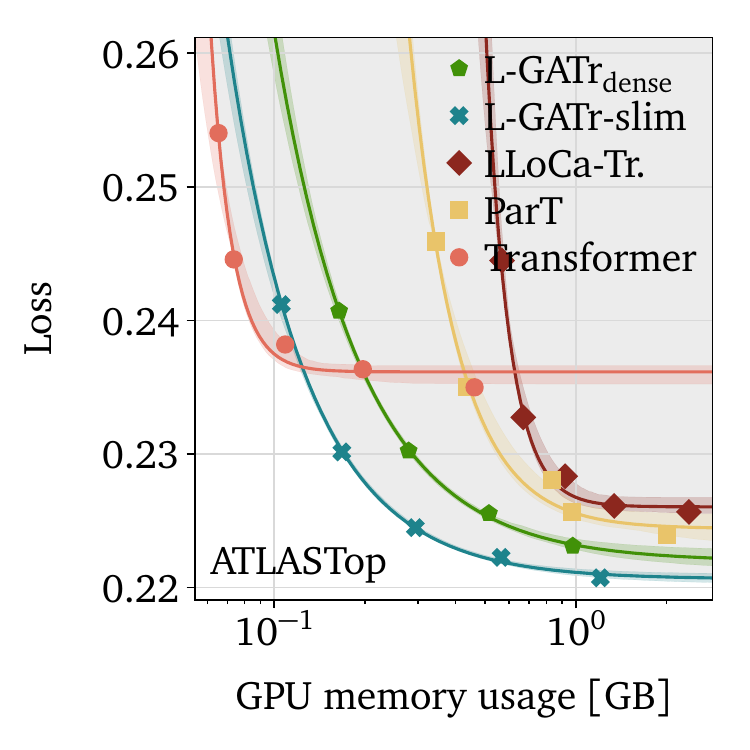}
    \includegraphics[width=0.325\linewidth,page=8]{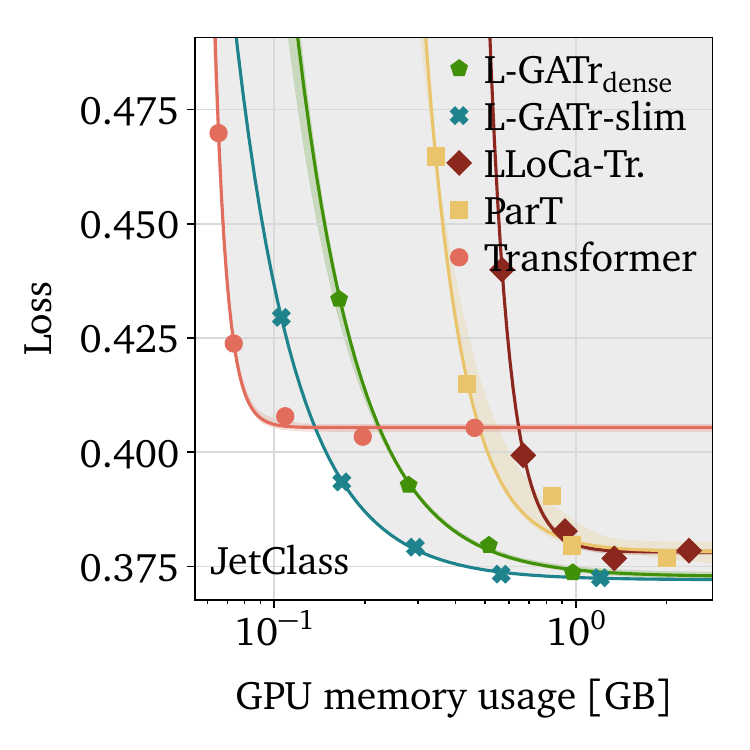}
    \includegraphics[width=0.325\linewidth,page=8]{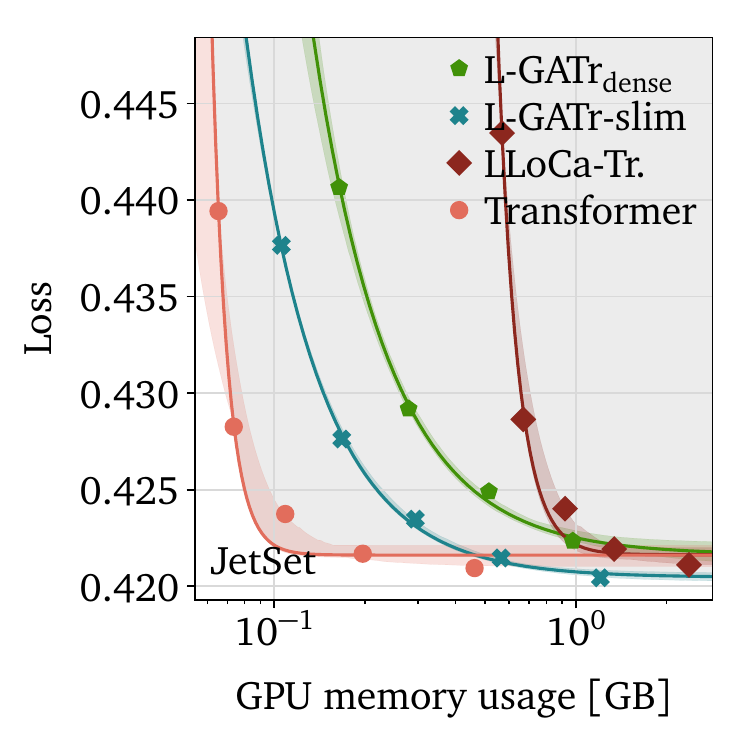}
    \caption{CPU (top) and GPU (bottom) memory use for the studied architectures. From left to right, we show the test loss over the inference metrics for ATLASTop, JetClass, and JetSet.}
    \label{fig:cost_3}
\end{figure}

We show the scaling fits for CPU and GPU memory use in Figure~\ref{fig:cost_3}. For all the datasets, the memory use on CPU is rather similar for architectures with the best performance-cost tradeoff on the studied scale. At around 0.1\,GB, the baseline transformer and L-GATr-slim perform similarly while at larger values L-GATr-slim wins by a large margin.

The increased memory use of L-GATr-slim and L-GATr at small network sizes is due to the additional memory required to store the vector and multivector channels. The LLoCa-Transformer shows competitive performance for intermediate network sizes, while falling slightly behind for larger networks due to larger loss values. The initial overhead observed both for the CPU and GPU memory use comes again from the frames predictor. Advanced network predictors might close this gap~\cite{Favaro:2025pgz}.

\section{Pre-training for top tagging}
\label{sec:pretraining}

\begin{table}[t]
    \centering
    \footnotesize
    \begin{tabular}{lllllrr} 
        \toprule
        Architecture & Accuracy & AUC & \multicolumn{2}{c}{$1/\epsilon_B$} & \#Para & \#Jets \\
        &&& $\epsilon_S=50\%$ & $\epsilon_S=30\%$ & & (Pre-)training \\
        \midrule
        ParticleNet \cite{Qu:2019gqs} & 0.940\dz{} & 0.9858 & \result{397}{\dz{}7} & \result{1615}{\dz{}93} & 0.4M & \multirow{8}{*}{1.2M~\cite{Kasieczka:2019dbj}} \\
        Transformer \cite{Favaro:2025pgz} & \result{0.9393}{0.0001} & 0.9855 & \result{389}{\dz{}6} & \result{1613}{118} & 2.0M \\
        ParT \cite{Qu:2022mxj} & 0.940\dz{} & 0.9858 & \result{413}{16} & \result{1602}{\dz{}81} & 2.1M \\
        LorentzNet* \cite{Gong:2022lye} & 0.942\dz{} & 0.9868 & \result{498}{18} & \result{2195}{173} & 0.2M \\
        PELICAN* \cite{Bogatskiy:2023nnw} & \result{0.9426}{0.0002} & 0.9870 & -- & \result{2250}{\dz{}75} & 0.2M \\
        LLoCa-Transformer* \cite{Favaro:2025pgz} & \result{0.9416}{0.0001} & 0.9866 & \result{492}{15} & \result{2150}{130} & 2.0M \\
        L-GATr* \cite{Brehmer:2024yqw} & \result{0.9423}{0.0002} & 0.9870 & \result{540}{20} & \result{2240}{\dz{}70} & 1.1M \\
        L-GATr-slim* \cite{Petitjean:2025zjf} & \result{0.9420}{0.0002} & 0.9869 & \result{546}{\dz{7}} & \result{2264}{\dz{}93} & 2.0M \\
        \midrule
        ParT-f.t. \cite{Qu:2022mxj} & 0.944\dz{} & 0.9877 & \result{691}{15} & \result{2766}{130} & 2.1M & \multirow{4}{*}{100M~\cite{Qu:2022mxj}} \\
        L-GATr-f.t.* \cite{Brehmer:2024yqw} & \result{0.9446}{0.0002} & 0.9879\dz{} & \result{651}{11} & \result{2894}{\dz{}84} & 1.1M \\
        L-GATr-slim-f.t.*\cite{Petitjean:2025zjf} & \result{0.9442}{0.0002} & 0.9879 & \result{655}{\dz{}5} & \result{2927}{\dz{}70} & 2.0M \\
        L-GATr-slim*-f.t. 48M \textbf{new} & \result{0.9446}{0.0001} & 0.9880 & \result{693}{17} & \result{3062}{\dz{}84} & 48M \\
        \midrule
        OmniLearned-M \cite{Bhimji:2025isp} & 0.944 & 0.9880 & \result{656}{12} & \result{3208}{176} & 58M & \multirow{2}{*}{1058M~\cite{Bhimji:2025isp}} \\
        OmniLearned-L \cite{Bhimji:2025isp} & 0.944 & 0.9880 & \result{688}{\dz{}9} & \result{3486}{157} & 423M \\
        \bottomrule
    \end{tabular}
    \caption{Top tagging accuracy, AUC and background rejection rates for two fixed signal efficiencies. Compared to earlier versions, the L-GATRs-slim ($s=2$) entry is new. We indicate Lorentz equivariance with an asterisk and estimate uncertainties using five trainings. Uncertainties in AUC are at most $\pm 0.0001$ and not reported.}
    \label{tab:top_tagging}
\end{table}

\begin{figure}[b!]
    \includegraphics[width=0.485\textwidth,page=3]{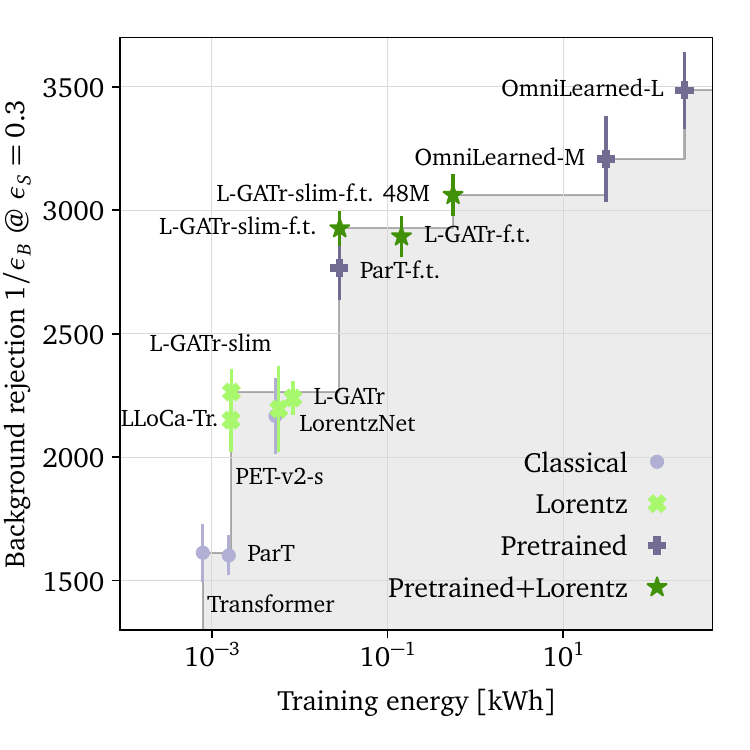} \hfill
    \includegraphics[width=0.485\textwidth,page=6]{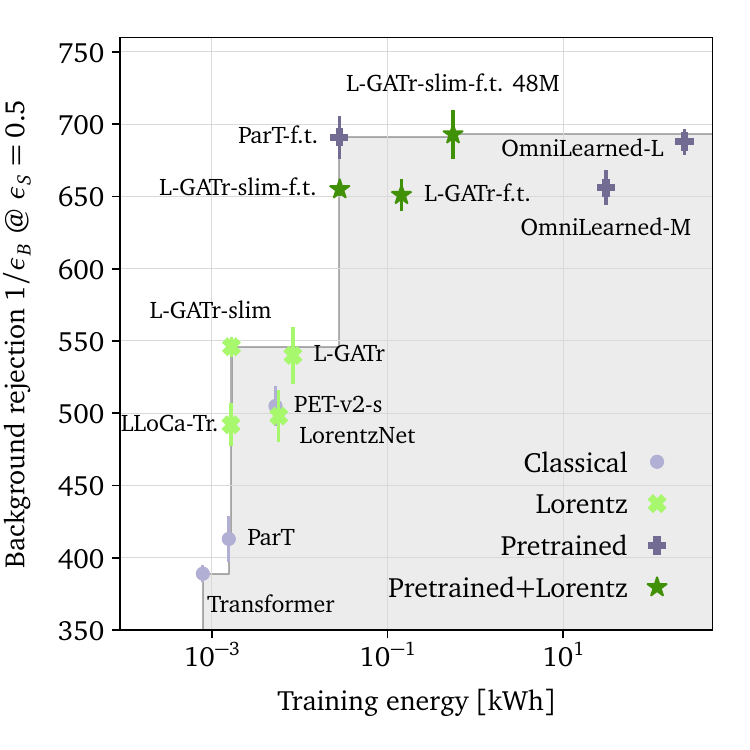}
    \caption{Background rejection rate at fixed signal efficiency $30\%$ (left) and $50\%$ (right) for top taggers over the estimated training energy use (right).}
    \label{fig:toptagging_energy}
\end{figure}

Finally, we use the representations learned by the large networks trained in the last sections and fine-tune a top tagger on the smaller, original benchmark dataset~\cite{Kasieczka:2019dbj}. 

We use L-GATr-slim as the leading equivariant architecture, as it gives the best performance at fixed cost, and we use the largest $s=2$ network with 48M parameters. We pretrain L-GATr-slim architecture only on the 4-momenta in the large JetClass dataset, using the hyperparameters from Section~\ref{sec:all_datasets}. The training needs around 33 hours on 4 H100 GPUs, summing to around 130 total GPU hours. The network is then finetuned on the smaller top tagging dataset, using the Lion optimizer, with learning rate $5\times 10^{-6}$ and weight decay $4$. 

\begin{figure}[t]
    \includegraphics[width=0.485\textwidth,page=1]{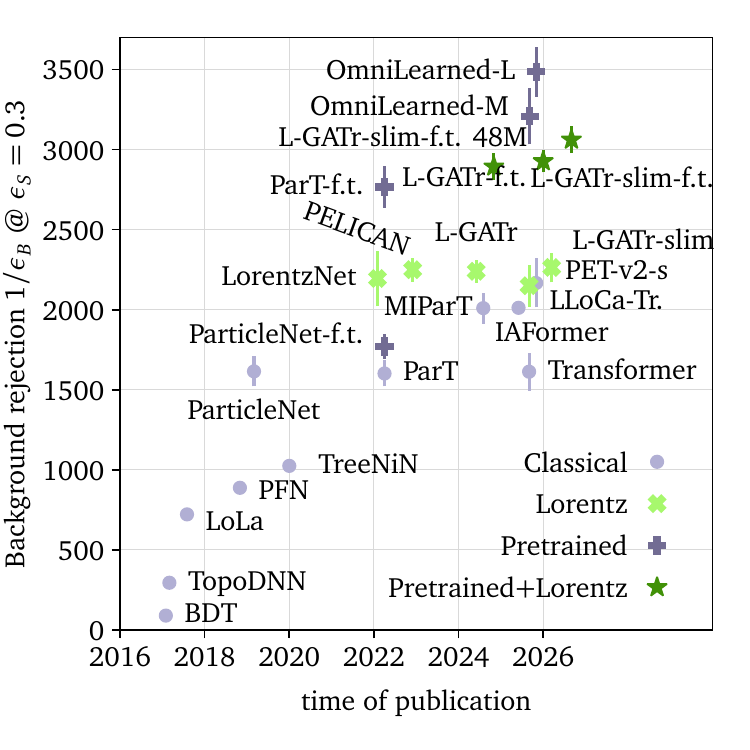} \hfill
    \includegraphics[width=0.485\textwidth,page=4]{figs/toptagging.pdf}
    \caption{Background rejection rate at fixed signal efficiency $\epsilon_s = 30\%$ (left) and $50\%$ (right) for top taggers over the time of publication.}
    \label{fig:toptagging_time}
\end{figure}

We summarize the results for from-scratch training and pre-training in Table~\ref{tab:top_tagging} and visualize them in Figure~\ref{fig:toptagging_time}. Pre-training uses either the 100M JetClass dataset or the 1B OmniLearned dataset\cite{Bhimji:2025isp}. We observe leading performance for ParT, L-GATr, and L-GATr-slim pre-trainings with parameter counts comparable to our $s=0$ networks, and the OmniLearned-M and OmniLearned-L architectures pretrained on the 1B dataset. The new implementation of L-GATr-slim $s=2$ shows that larger networks benefit from pre-training. It matches the performance of the larger OmniLearned architectures in accuracy, AUC, and background rejection rate at fixed signal efficiency 0.5, and only performs worse on the background rejection rate at $30\%$ signal efficiency.

Following the theme of the paper, we also show the energy use of these trainings in Figure~\ref{fig:toptagging_energy}. We estimate the energy use per forward pass as in  Section~\ref{sec:tagging_networks}. We perform the calculation for a backward pass and then multiply by the number of iterations and batch size for pre-training and fine-tuning. We include an additional factor for the average number of particles in a batch which favors networks using sparse representations instead of zero-padding. We visualize the Pareto frontier of highest performance at given training energy use. The most efficient architectures for small budgets are the transformer and L-GATr-slim, the latter outperforming PET-v2-S and LorentzNet. At larger training energy consumption, L-GATr-slim with 1M and 48M parameters, pre-trained on the JetClass dataset, are optimal. Finally, OmniLearned-M and OmniLearned-L require the largest compute budgets.

The ranking of the models changes with the evaluation metric, for instance, the background rejection rates at signal efficiencies $30\%$ and $50\%$. At the largest rejection rates, the fine-tuning  hyperparameters appear to affect which quantity the network focuses on. We find that OmniLearned-L with 300M parameters and pre-trained on the 1B dataset outperforms ParT-f.t. with 2M parameters, pre-trained on the 100M dataset significantly at signal efficiency $30\%$. In contrast, they perform similarly at signal efficiency $50\%$. Also, the top tagging evaluation dataset is relatively small with 400k jets, making it difficult to distinguish between the most advanced taggers.

\section{Outlook}
\label{sec:outlook}

Jet tagging is the flagship ML application in LHC physics. In practice, its concept and tool development is split into large-$R$ jet tagging and flavor tagging. For the latter, Lorentz-equivariant transformers lead in performance, raising the question if flavor taggers can benefit from the same symmetry. Such a synergy it suggested by the development of foundation models for the LHC, where we assume that the network performance benefits from learned, common underlying structures and symmetries. 

To realize such synergies we developed unified taggers for large-$R$ jets and flavor jets, building on physics-inspired improvements over a baseline transformer. The obvious physics structure to use at the LHC is Lorentz-invariance of relativistic particle kinematics. The difference between explicit and implicit symmetry implementations is softened by the fact that strictly equivariant transformers have to learn experimental symmetry breaking patterns. The real question is if we want to learn broken symmetries directly or as the breaking of explicit symmetries. 

As a starting point, we provided efficient implementations of baseline, physics-inspired, and Lorentz-equivariant transformers and compared different evaluation cost metrics. We first benchmarked the raw tagging performance on an ATLAS dataset with realistic top jets, the JetClass dataset with many different large-$R$ jets, and an ATLAS dataset used to train flavor taggers. We found that Lorentz-equivariant architectures win whenever 4-vectors are relevant, and match the baseline performance when scalar information is important. 

Next, we determined the performance at fixed evaluation cost, like energy consumption, inference time, and memory. For very low cost the baseline transformer performed best, but above a very basic threshold equivariant transformers outperform (ATLASTop/JetClass) or at least match it (JetSet). The leading Lorentz-equivariant architecture throughout was L-GATr-slim~\cite{Brehmer:2024yqw,Petitjean:2025zjf}.

From a representation learning and foundation model perspective, all studied transformers can and should be used across LHC physics, as the underlying structures and symmetries are universal. The fact that flavor tagging, building on scalar input, and large-$R$ tagging, building on Lorentz-invariants, behave so differently raises the interesting question what kind of symmetry or structure we expect such foundation models to encode or learn.

\subsection*{Code availability}

Our code is available at \url{https://github.com/heidelberg-hepml/tagging-guide}. The efficient implementation of all architectures included in this paper are available in the L-GATr package (L-GATr and L-GATr-slim) at \url{https://github.com/heidelberg-hepml/lgatr}, and the LLoCa package (Transformer, ParT, LLoCa-Transformer) at \url{https://github.com/heidelberg-hepml/lloca}. The ParT implementation in the LLoCa package is based on the public implementation available at \url{https://github.com/hqucms/weaver-core}.

\section*{Acknowledgements}

We thank Víctor Bréso-Pla and Antoine Petitjean for their feedback on the improved L-GATr and L-GATr-slim implementation. We also thank Andrea Giammanco, Matthias Vigl, Nicole Hartman, Michael Kagan, and Lukas Heinrich for useful discussions, and Daniel Whiteson and Ben Nachman for pushing to make the critical ATLASTop dataset public. L.F. is supported by the Fonds de la Recherche Scientifique - FNRS under Grant No. 4.4503.16. T.P. is supported by the Deutsche Forschungsgemeinschaft (DFG, German Research Foundation) under grant 396021762 – TRR 257 Particle Physics Phenomenology after the Higgs Discovery.  J.S. gratefully acknowledges support from the Alexander-von-Humboldt foundation as a Feodor Lynen Fellow. Computational resources have been provided by the supercomputing facilities of the Université catholique de Louvain (CISM/UCL) and the Consortium des Équipements de Calcul Intensif en Fédération Wallonie Bruxelles (CÉCI) funded by the Fond de la Recherche Scientifique de Belgique (F.R.S.-FNRS) under convention 2.5020.11 and by the Walloon Region. The presented research benefited from computational resources made available on Lucia, the Tier-1 supercomputer of the Walloon Region, infrastructure funded by the Walloon Region under the grant agreement n°1910247. The authors acknowledge support by the state of Baden-Württemberg through bwHPC and the German Research Foundation (DFG) through the grants INST 35/1597-1 FUGG and INST 39/1232-1 FUGG. The authors gratefully acknowledge the computing time provided on the high-performance computer HoreKa by the National High-Performance Computing Center at KIT (NHR@KIT). This center is jointly supported by the Federal Ministry of Education and Research and the Ministry of Science, Research and the Arts of Baden-Württemberg, as part of the National High-Performance Computing (NHR) joint funding program (https://www.nhr-verein.de/en/our-partners). HoreKa is partly funded by the German Research Foundation (DFG).

\clearpage
\appendix
\section{Variance preservation in LLoCa}
\label{app:variance}

This appendix describes an improvement in the Lorentz Local Canonicalization (LLoCa)~\cite{Favaro:2025pgz} framework that improves training stability in particular for highly boosted objects. 
The critical operations are the frame-to-frame transformations in the attention operation. We recall the LLoCa attention from Ref.~\cite{Favaro:2025pgz},
\begin{align}
    f_{i,L_i}^\text{updated} = \rho(L_i)\sum_{j=1}^N \text{softmax}\Big(\frac{1}{\sqrt{d}}\left\langle \rho (L_i^{-1})q_{i,L_i},\rho_k(L_j^{-1})k_{j,L_j}\right\rangle\Big) \rho(L_j^{-1})v_{j,L_j}\;.
\end{align}
In this expression, $\rho (L_i)$ denotes the representation of the frame transformation matrix $L_i$, and choosing the trivial representation restores the standard scaled dot-product attention. However, for the vector representation, $\rho (L_i)=L_i$ acts as a $4\times 4$ matrix. If this matrix $L_i$ has large entries, its inverse $L_i^{-1}$ will also have large entries, and they jointly drive up the scale of the latent representation across layers, causing unstable training.

Deep neural networks achieve stable training by keeping the variance of their latent representation constant across the network layers. The widely used Xavier-Glorot~\cite{Glorot:2010init} and Kaiming-He~\cite{He:2015rectifiers} weight initialization schemes implement this idea. To apply the same concept to the frame transformations in LLoCa, we consider random latent vectors $x\in\mathbb{R}^4$ and evaluate their covariance matrix $\Sigma = \mathbb{E}_x[x x^T]$, as well as the total variance $\text{tr}(\Sigma)$. For uncorrelated latent vectors, we have $\Sigma = \sigma^2 \mathbb{1}$ and total variance $4\sigma^2$. Applying a fixed Lorentz transformation $L\in\mathbb{R}^{4\times 4}$, we write the transformed variance as
\begin{align}
    \text{tr}\left(\mathbb{E}_x\left[(Lx)(Lx)^T) \right]\right) = \text{tr}\left(L \mathbb{E}_x[x x^T]L^T\right) = \sigma^2\text{tr}\left(L L^T\right) = \sigma^2\sum_{i,j=1}^4 L_{ij}^2 \equiv \sigma^2 \|L\|_F^2\;.
\end{align}
The variance has increased by a factor of $\|L\|_F^2/4$ relative to the original variance $4\sigma^2$, where $\|L\|_F$ is the Frobenius norm of $L$. To preserve the variance of the latent representation $x$ under $x\to Lx$, we have to apply the Lorentz transformation $2L/\|L\|_F$. Note that pure rotations have $L L^T=1_4$ such that $\|L\|_F^2=4$, and therefore no scaling is required, while for a Lorentz transformation with boost factor $\gamma$,  $\|L\|_F = 2\gamma$. 

To apply the previous discussion to LLoCa, $\gamma$ has to be extracted from a Lorentz-invariant frame transformation matrix.
A simple choice is to use a Lorentz-invariant canonicalization of the data, i.e. to boost the network inputs to the jet rest frame, such that the lab frame and relative boost factor $\gamma$ are invariant.
The more flexible approach is to keep a non-invariant lab frame, but still use the boost factor $\gamma$ relative to an invariant frame defined by a reference Lorentz vector $p_\text{ref}$ as $\gamma = (L p_\text{ref})^0 / \sqrt{\langle p_\text{ref},p_\text{ref}\rangle}$, where $(Lp_\text{ref})^0$ denotes the time component of the transformed vector $Lp_\text{ref}$. The rest frame corresponds to a sum over all momenta, $p_\text{ref}=\sum_i p_i$, and the lab frame is recovered with the non-invariant choice $p_\text{ref}=(1,0,0,0)$.
More generally, for a tensor of order $n$, the scaling factor becomes $1/\gamma^n$, or, equivalently, we can replace $L\to L/\gamma$. Restoring particle indices in the scaling factors $1/\gamma_i$, we obtain the refined LLoCa attention mechanism
\begin{align}
    f_{i,L_i}^\text{updated} = \rho(L_i/\gamma_i)\sum_{j=1}^N \text{softmax}\Big(\frac{1}{\sqrt{d}}\left\langle \rho (L_i^{-1}/\gamma_i)q_{i,L_i},\rho_k(L_j^{-1}/\gamma_j)k_{j,L_j}\right\rangle\Big) \rho(L_j^{-1}/\gamma_j)v_{j,L_j}\;,
\end{align}
which we use in all our trainings. We use the jet momentum $p_\text{jet}$ to define the invariant global frame, leading to
\begin{align}
    \gamma_i = \frac{(L_i p_\text{jet})^0}{m_\text{jet}},\quad m_\text{jet}^2 = \langle p_\text{jet},p_\text{jet}\rangle\;.
\end{align}

\begin{figure}
    \centering
    \includegraphics[width=0.325\linewidth,page=1]{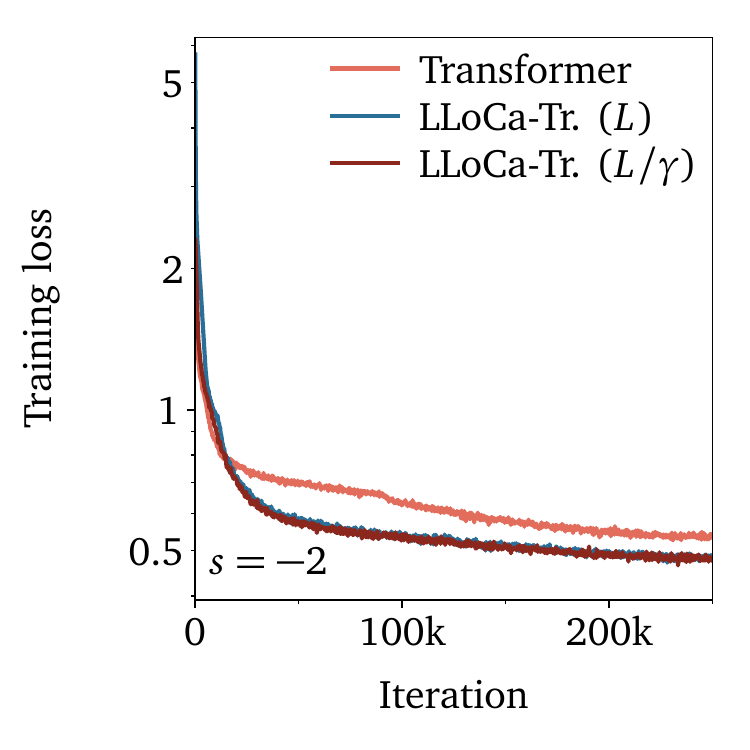}
    \includegraphics[width=0.325\linewidth,page=3]{figs/tracking.pdf}
    \includegraphics[width=0.325\linewidth,page=5]{figs/tracking.pdf}
    \includegraphics[width=0.325\linewidth,page=2]{figs/tracking.pdf}
    \includegraphics[width=0.325\linewidth,page=4]{figs/tracking.pdf}
    \includegraphics[width=0.325\linewidth,page=6]{figs/tracking.pdf}
    \caption{Loss (left), gradient norm (middle) and logits variance (right) of Transformer and LLoCa-Transformer taggers ($s=-2$ top and $s=2$ bottom) trained on JetClass for 250k iterations with batchsize 128 and following the default hyperparameters of Table~\ref{tab:arch-scaling} otherwise and Section~\ref{sec:all_datasets} otherwise.}
    \label{fig:preserve_variance}
\end{figure}

We visualize the impact of this choice in Figure~\ref{fig:preserve_variance}, where we show training curves for LLoCa-Transformer networks with $L/\gamma$ compared to $L$ only, and to a baseline transformer. For $s=-2$ (40k network parameters, 4 blocks) all architectures behave similarly, whereas for $s=2$ (50M network parameters, 17 blocks) the LLoCa-Transformer without variance preservation $L/\gamma$ is significantly less stable, while the refined LLoCa-Transformer behaves similarly to the baseline transformer. 
The effect of the variance preservation term becomes important at large network sizes because the problematic increase in the vector norm accumulates over transformer blocks.
%

\section{Additional results}
\subsection*{Compute cost results}
\label{app:cost}

\begin{figure}[tb]
\centering
    \includegraphics[width=0.325\linewidth,page=8]{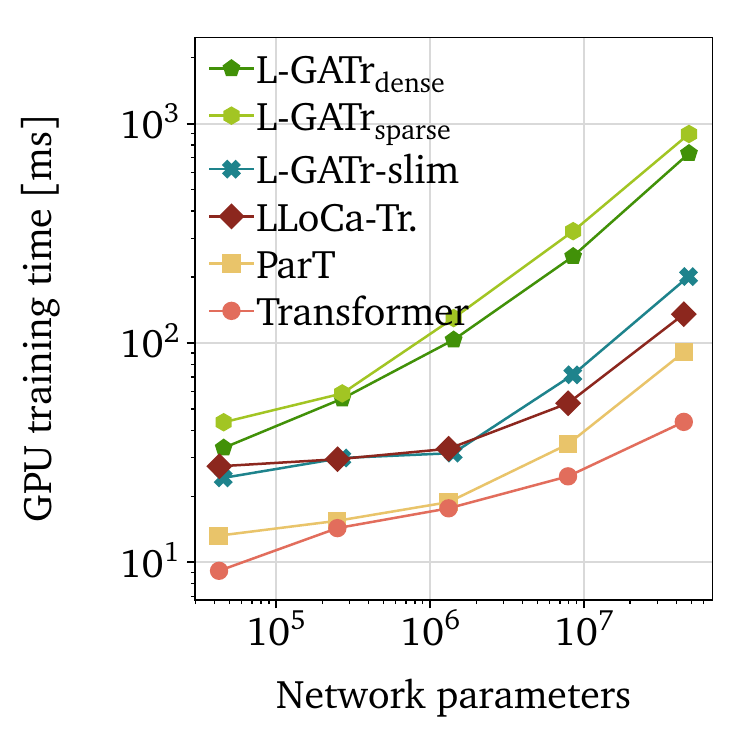}
    \includegraphics[width=0.325\linewidth,page=9]{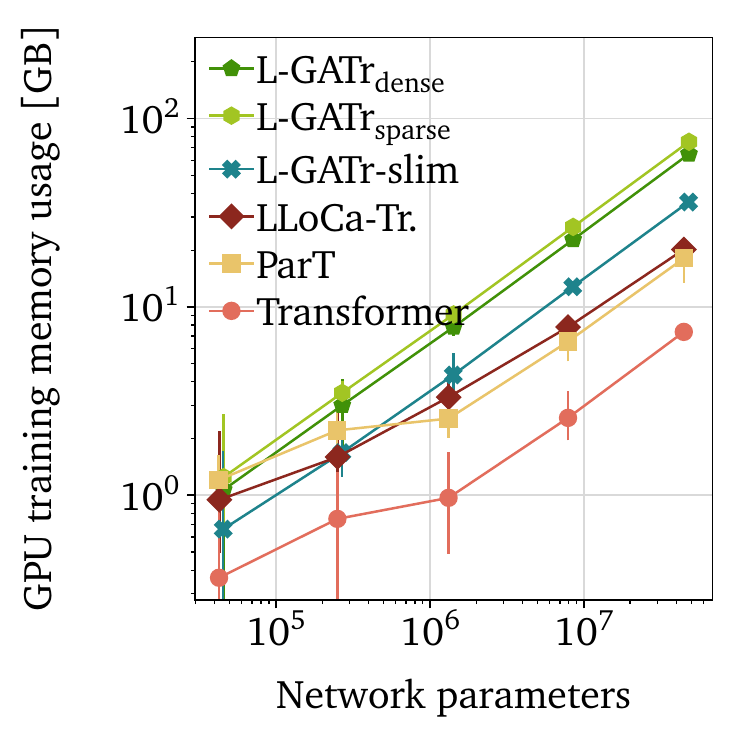}
    \caption{Training time (left) and memory use (right) scaling of the different architectures. }
    \label{fig:cost_train}
\end{figure}

We show training time per iteration and memory use scaling in Figure~\ref{fig:cost_train}, similar to the scaling of inference cost metrics in Figure~\ref{fig:cost}. Compared to the GPU inference measurements reported in Figure~\ref{fig:cost}, times are roughly $3\times$ larger due to the backward pass. The memory consumption is significantly larger, because activations have to be stored for the backward pass.

\begin{figure}[tb]
    \includegraphics[width=0.325\linewidth,page=8]{figs/cost_gains.pdf}
    \includegraphics[width=0.325\linewidth,page=9]{figs/cost_gains.pdf}
    \includegraphics[width=0.325\linewidth,page=10]{figs/cost_gains.pdf} \\
    \includegraphics[width=0.325\linewidth,page=1]{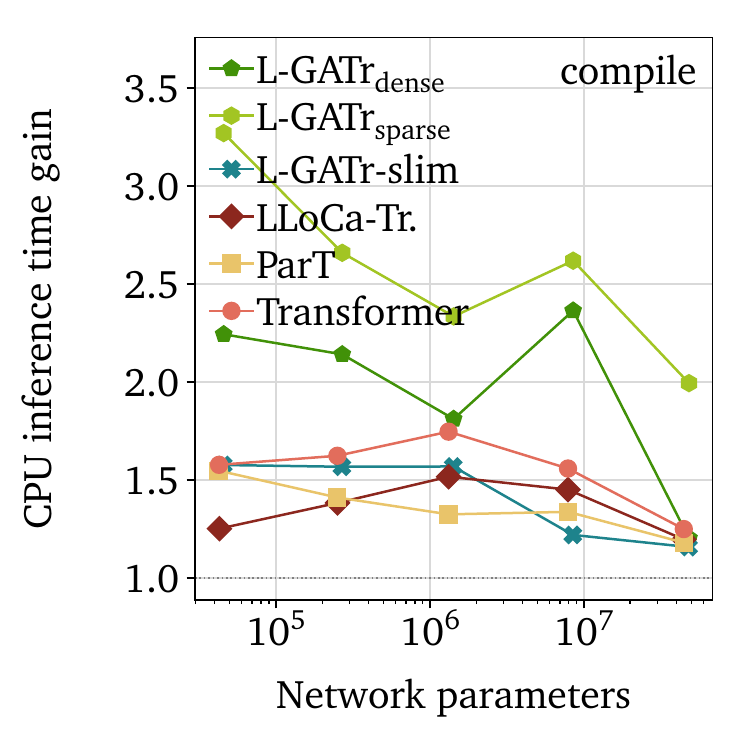}
    \includegraphics[width=0.325\linewidth,page=2]{figs/cost_gains.pdf}
    \caption{Relative gains on GPU inference memory (top row) and CPU inference time(bottom row) from \texttt{torch.compile}, AMP, and sparse jet representations. }
    \label{fig:cost_gains2}
\end{figure}

In Figure~\ref{fig:cost_gains2} we show the relative gains from using \texttt{torch.compile}, AMP, and sparse jet representations on the GPU inference memory use and CPU inference time, similar to the gains for GPU inference time discussed in Figure~\ref{fig:cost_gains}. 

\clearpage
\subsection*{Resilience}
\label{app:resilience}

Equivariant neural networks are often motivated by two complementary considerations. First, when implemented efficiently, they can achieve better performance for a fixed computational budget by incorporating known symmetries directly into the model architecture. Second, the constraints imposed by these symmetries reduce the space of allowed functions, which may influence how the network responds to variations that are not explicitly encoded in the training data. However, whether this architectural bias translates into improved resilience to systematic effects is ultimately an empirical question.

This question is particularly relevant for jet tagging, where increasingly expressive architectures operating on low-level detector information have achieved substantial gains in discrimination power, but often at the cost of a stronger dependence on the details of simulated training samples. As a consequence, systematic uncertainties associated with event generators, detector modeling, and experimental conditions become increasingly important~\cite{ATLAS:2024rua}. Although many experimental analyses mitigate simulation mis-modeling through calibration procedures and data-driven corrections, residual systematic uncertainties remain an important limitation. Studying the intrinsic sensitivity of taggers to these variations is, therefore, still valuable, as it determines the size of the corrections required and the robustness of the classifier when applied to data.

In this section, we investigate whether imposing Lorentz equivariance can improve tagging performance while maintaining, or potentially reducing, the sensitivity to these systematic variations. To this end, we use the dataset and uncertainty definitions introduced in~\cite{ATLAS:2024rua} to assess the resilience of Lorentz-equivariant jet taggers.

\paragraph{Modeling uncertainties}

Starting from the theoretical uncertainties, we evaluate the taggers on pairs of datatsets produced with different simulators. Following the definition used in~\cite{ATLAS:2024rua}, we define the following systematics:
\begin{itemize}
    \item signal modeling: from top jets produced in $t\bar{t}$ using Pythia8 or Herwig7;
    \item background modeling: this includes parton shower effects, using two samples generated with either the angular ordering or the dipole modeling in Herwig, and the hadronization, from two more samples produced using either the cluster model or the Lund string model in Sherpa.
    \item scale uncertainty: defined as factors of two scale variations in the Pythia8 shower weights. These variations cover the effect of renormalization and factorization scales used for the intial and final state radiation, and for the background and signal samples.
\end{itemize}
The varied top samples are generated from the Powheg Box v2 event generator at NLO with the $\text{NNPDF3.0}_\text{NLO}$ PDF set and the $h_\text{damp}$ parameter set to 1.5 times the mass of the top quark. The other background samples are generated at LO with the $\text{NNPDF3.0}_\text{LO}$ PDF set~\cite{ATLAS:2024rua}.\medskip

\begin{figure}[t]
    \includegraphics[width=0.325\linewidth,page=2]{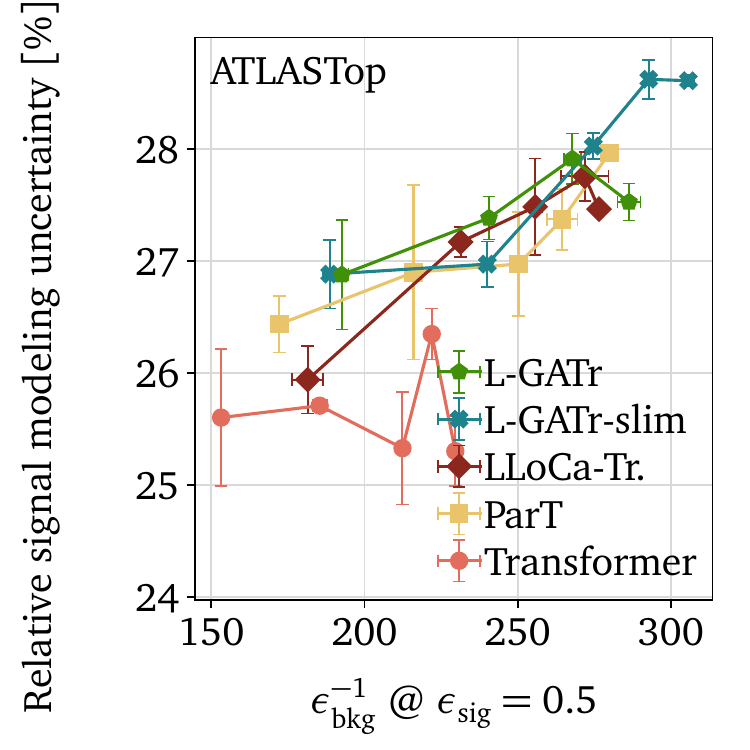}
    \includegraphics[width=0.325\linewidth,page=4]{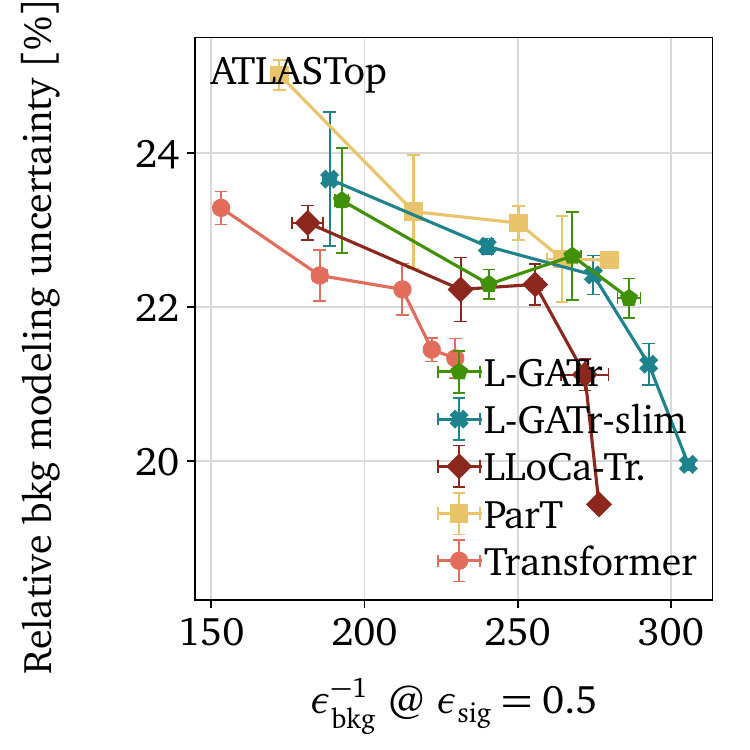}
    \includegraphics[width=0.325\linewidth,page=3]{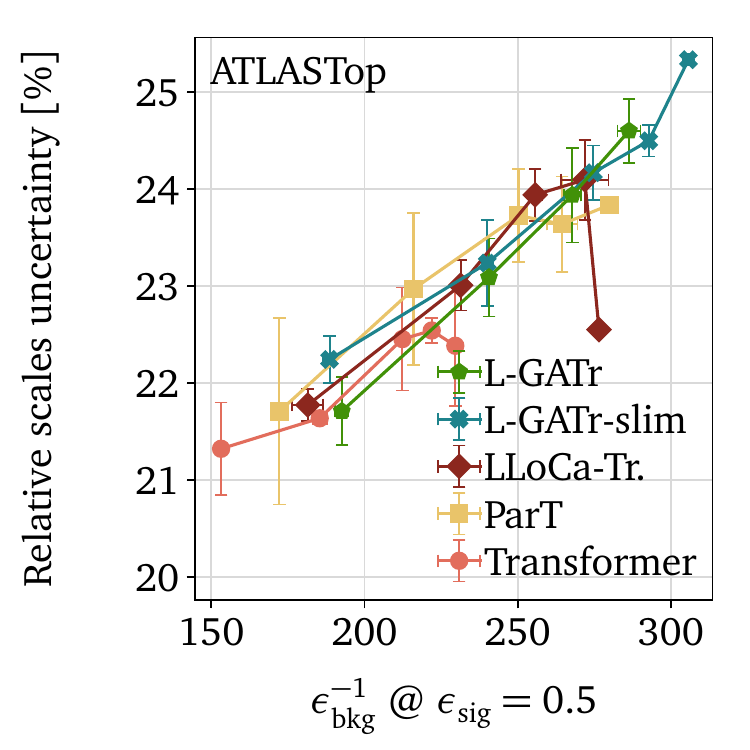}
    \caption{Theoretical uncertainties for the ATLAS top tagging dataset. From left to right, we show $\sigma_\text{sig}$, $\sigma_\text{bkg}$, and $\sigma_\text{scale}$ for taggers with different sizes and architectures.}
    \label{fig:unc_atlastop_th}
\end{figure}

We evaluate the background rejection at signal efficiency of $50\%$ for each pair of datasets and calculate the uncertainty as the relative difference between the two datasets, i.e.
\begin{align}
    \sigma_\text{th} = \frac{\epsilon_\text{bkg,1}^{-1}(0.5)-\epsilon_\text{bkg,2}^{-1}(0.5)}{\epsilon_\text{bkg,1}^{-1}(0.5)} \,.
    \label{eq:rel_unc}
\end{align}
We sum in quadrature the two background uncertainties to define a single $\sigma_\text{bkg}$, and, similarly, we sum the four scale uncertainties into $\sigma_\text{scale}$.
Figure~\ref{fig:unc_atlastop_th} shows that taggers with embedded symmetries, L-GATr, L-GATr-slim, LLoCa-Tr., and ParT, have improved performance at modest increase in modeling uncertainty. In particular, the signal modeling shows an improvement of a factor two without any increase in uncertainty. We estimate the error on each network uncertainty by retraining the tagger three times with different initializations.

\paragraph{Experimental uncertainties}

The experimental uncertainties are divided into
\begin{itemize}
    \item track uncertainties: these are divided into track efficiency, fake rate, and bias systematic variations. These are uncertainty estimates for the efficiency in finding the true tracks, the fake rate from alignments, and residual bias in the measurement of the transverse momentum.
    \item cluster uncertainties: these are divided into cluster energy scale, energy resolution, and position systematic variations. These are uncertainty estimates for the energy response of the ATLAS calorimeter, its energy and position resolution.
\end{itemize}
A detailed discussion on the generation of these samples can be found in Ref.~\cite{ATLAS:2024rua}. \medskip

We calculate the cluster energy and position resolution, and the track bias uncertainties similarly to Eq.\ref{eq:rel_unc}, by taking the relative difference of the background efficiency between the nominal samples and the varied one. For the cluster energy scale, we instead take the envelope between a systematic variation up or down of the cluster energy. In equations, 
\begin{align}
\sigma_\text{cluster\, scale} 
    &= \max(\sigma_\text{cluster\, up}, \sigma_\text{cluster\, down}) \,, \\
\sigma_\text{cluster\, up/down} 
    &= \frac{\epsilon_\text{bkg,nom.}^{-1}(0.5)-\epsilon_\text{bkg,up/down}^{-1}(0.5)}{\epsilon_\text{bkg,nom.}^{-1}(0.5)} \,.
\end{align}
We use the same formula for the remaining track uncertainties, which include variations at the global level of charged tracks or merged tracks. Therefore, these are described by
\begin{align}
\sigma_\text{track\, syst} 
&= \max(\sigma_\text{track\, global}, \sigma_\text{track\, jet})\,, \\
\sigma_\text{track\, global/jet} 
&= \frac{\epsilon_\text{bkg,nom.}^{-1}(0.5)-\epsilon_\text{bkg,global/jet}^{-1}(0.5)}{\epsilon_\text{bkg,nom.}^{-1}(0.5)} \,.
\end{align}
where $\sigma_\text{track\, syst}$ indicates either the track fake rate or the track efficiency. \medskip

\begin{figure}[t]
    \includegraphics[width=0.325\linewidth,page=10]{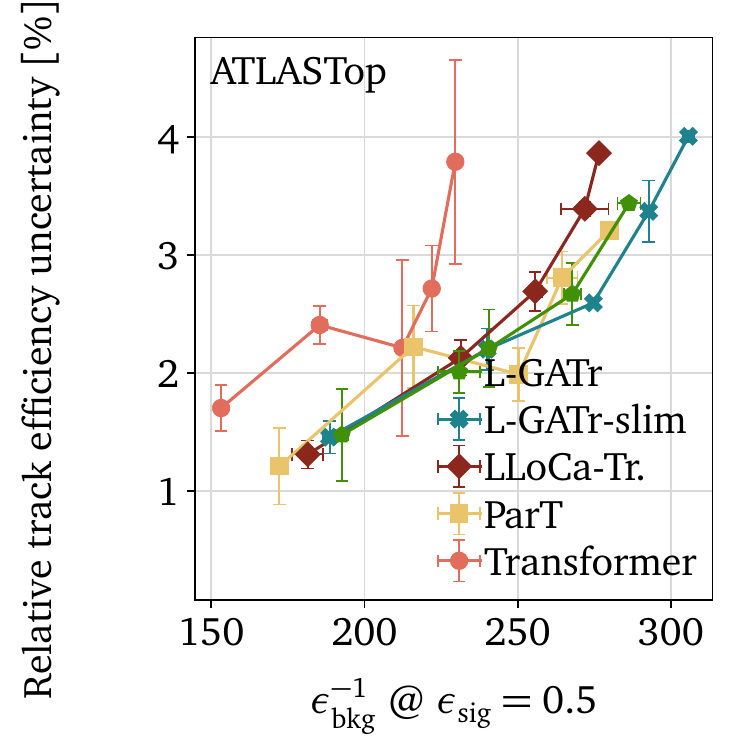}
    \includegraphics[width=0.325\linewidth,page=11]{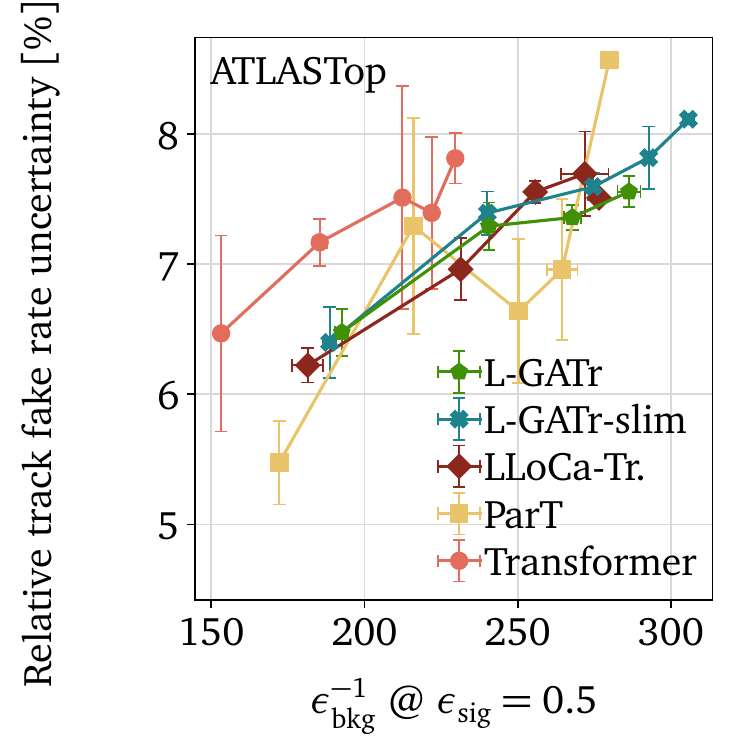}
    \includegraphics[width=0.325\linewidth,page=5]{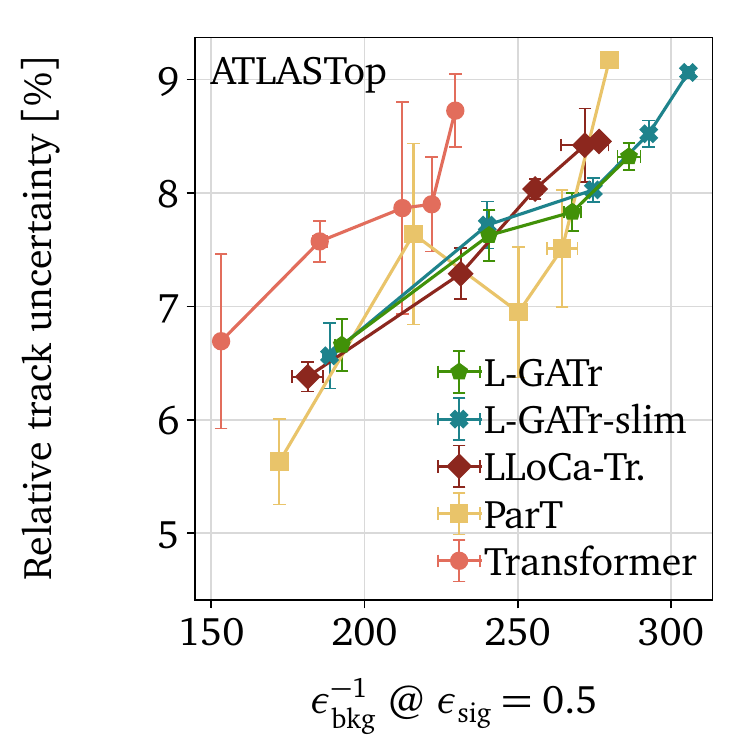} \\
    \includegraphics[width=0.325\linewidth,page=7]{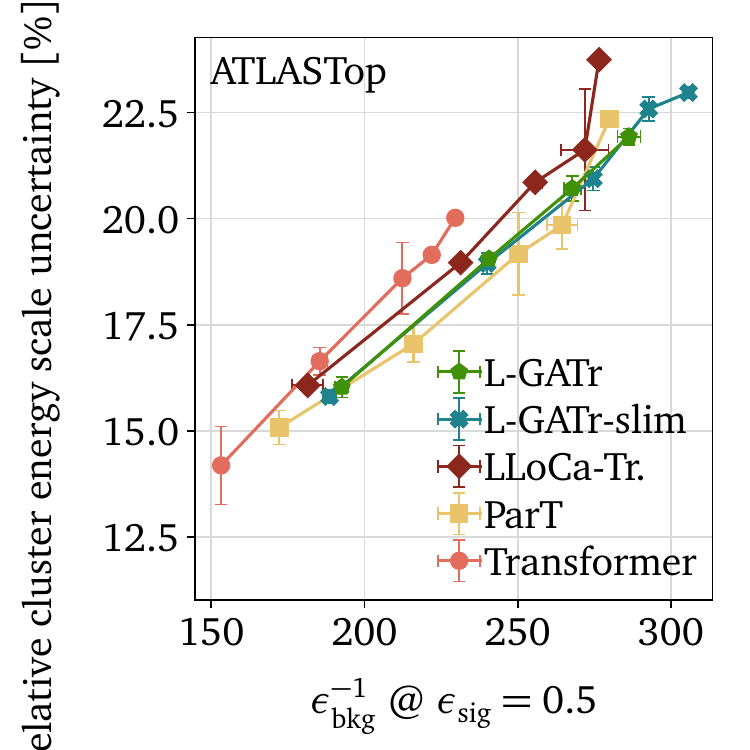}
    \includegraphics[width=0.325\linewidth,page=8]{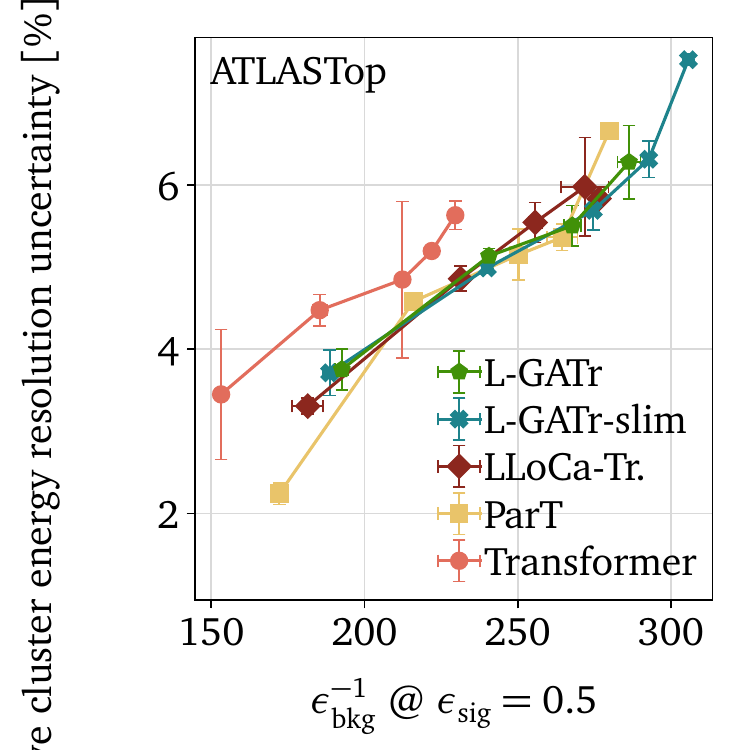}
    \includegraphics[width=0.325\linewidth,page=6]{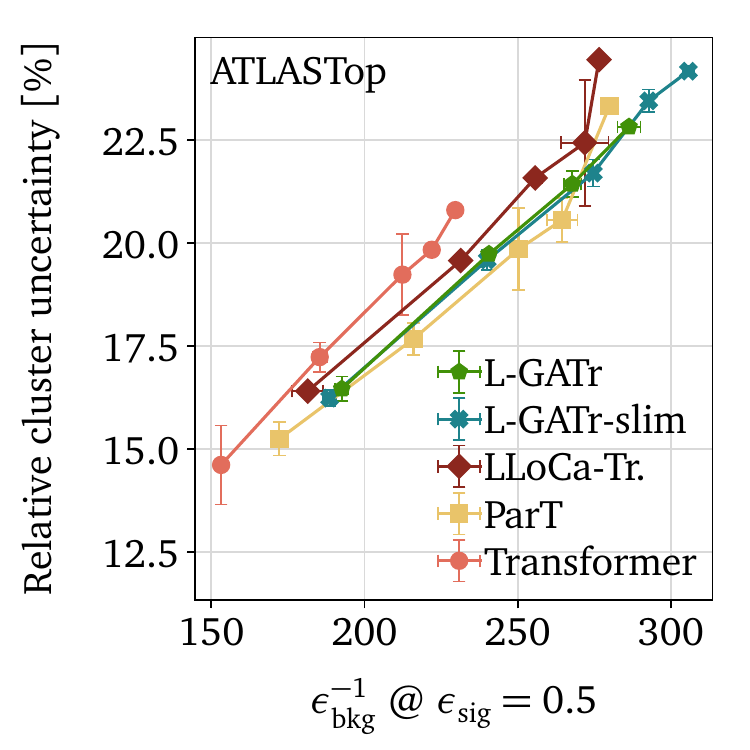}
    \caption{Experimental uncertainties for the ATLAS top tagging dataset. (top) From left to right, we show $\sigma_\text{track\,eff.}$, $\sigma_\text{track\,fake}$, and the total $\sigma_\text{track}$ for taggers with different sizes and architectures. (bottom) Similarly, we show $\sigma_\text{cluster\,scale}$, $\sigma_\text{cluster\,res.}$, and the total $\sigma_\text{cluster}$.}
    \label{fig:unc_atlastop_exp}
\end{figure}

Figure~\ref{fig:unc_atlastop_exp} shows, in each row, the two main uncertainties for tracking and clustering for the considered taggers. In the last column, we show the total uncertainty calculated as the squared sum of the others. Although we observe a noisy evaluation due to the small size of the samples, the symmetry-aware netwroks show an improved performance without an increase in the systematic uncertainties. This is represented by a shift towards larger background rejections at fixed network size.

\begin{figure}[t]
    \includegraphics[width=0.325\linewidth,page=19]{figs/unc_atlastop5ep_final.pdf}
    \includegraphics[width=0.325\linewidth,page=20]{figs/unc_atlastop5ep_final.pdf}
    \includegraphics[width=0.325\linewidth,page=1]{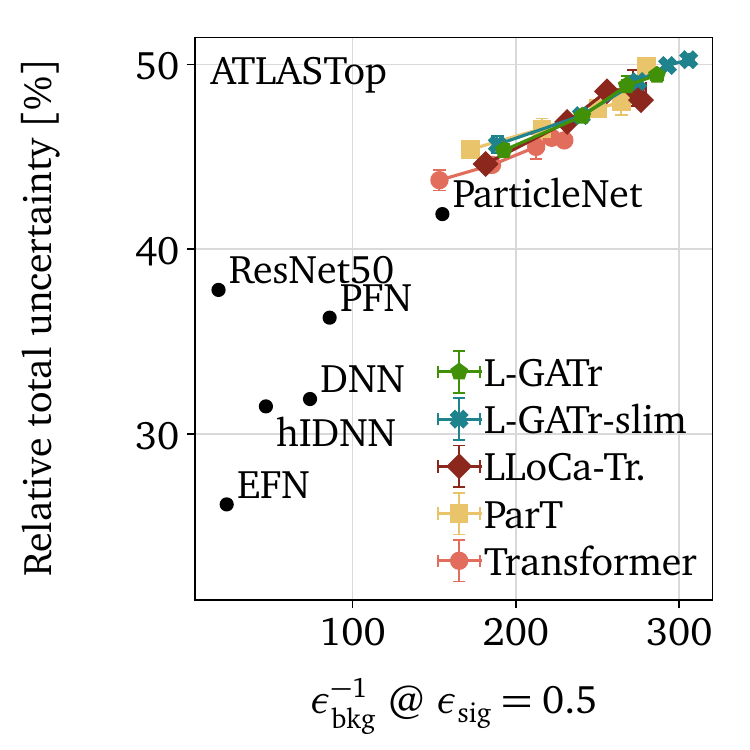}
    \caption{Summary of (left) experimental, (middle) modeling, and (right) total relative uncertainties for the ATLAS top tagging dataset.}
    \label{fig:unc_atlastop_summary}
\end{figure}

Finally, we summarize the evaluation in total experimental and modeling uncertainties, as shown in Figure~\ref{fig:unc_atlastop_summary}. These are calculated as the quadrature sum following the classification of the previous paragraphs. Additionally, we also provide the total relative uncertainty together with the networks already evaluated in Ref.~\cite{ATLAS:2024rua}.
We observe that Lorentz-equivariant networks increase the performance slightly affecting the experimental uncertainties. On the other hand, the modeling uncertainties increase by a few percents as we see the Lorentz-equivariant network above the standard transformer. Rather surprisingly, the LLoCa-Transformer and the standard transformer have a sharp drop in modeling uncertainties at $s=2$ which will require further study. The sum of the two effects results in a total relative uncertainty which lies on a similar diagonal already identified in Ref.~\cite{ATLAS:2024rua}. We note that this definition of uncertainties does not take into account the inherently different optimal performance for the various samples as it would be possible with a conditional training~\cite{Butter:2022xyj}. A more faithful evaluation of the network uncertainties should use such a conditional training which, however, is not possible due to the limited size of the samples. We leave this exploration to future work.

\subsection*{Salt/GN results}
\label{app:gn3}

In this section, we show the scaling results for the transformers trained using the Salt repository~\cite{salt2025}, which is the main reference for the training of the GN networks by the ATLAS collaboration. To obtain these results, we minimally modified the Salt repository to make it compatible with our codebase and trained networks using the same scaling defined in Sec.~\ref{sec:tagging_networks}. In addition to the usual scaling of the number of transformer channels, blocks, and heads, we scale the final three fully-connected linear layers by an increasing factor of two, as shown in Table~\ref{tab:salt_arch_scaling}. Besides the additional three output layers, the Salt transformer uses learnable tokens in the transformer layer, SiLU nonlinearities, and a learnable weighted aggregation. We do not show the cost metrics for these networks because we cannot guarantee with the same standards that the code is optimized consistently with the other networks.

\begin{table}[t]
\centering
    \begin{small} \begin{tabular}{ll rrrrrr}
    \toprule
    & & XS & S & M & L & XL & \\
    & & $s=-2$ & $s=-1$ & $s=0$ & $s=1$ & $s=2$ & $s$ \\
    \midrule
    Channels & & 32 & 64 & 128 & 256 & 512 & $2^{7+s}$ \\
    Blocks & & 4 & 6 & 8 & 12 & 17 & $3\cdot 2^{\lfloor (s+3)/2\rfloor}$ \\
    Heads & & 4 & 4 & 8 & 8 & 16 & $2^{3 + \lfloor s/2\rfloor}$\\
    \midrule
    \multirow{3}{*}{Salt/GN}
    & output layer 1 & 32 & 64 & 128 & 256 & 512 & $2^{7+s}$ \\
    & output layer 2 & 16 & 32 & 64 & 128 & 256 & $2^{6+s}$ \\
    & output layer 3 & 8 & 16 & 32 & 64 & 128 & $2^{5+s}$ \\
    \bottomrule
    \end{tabular} \end{small}
    \caption{Scaling prescription for the Salt/GN transformer. We scale the last three network layers by an increasing factor of two.}
    \label{tab:salt_arch_scaling}
\end{table}

\begin{figure}[t]
    \includegraphics[width=0.485\linewidth,page=1]{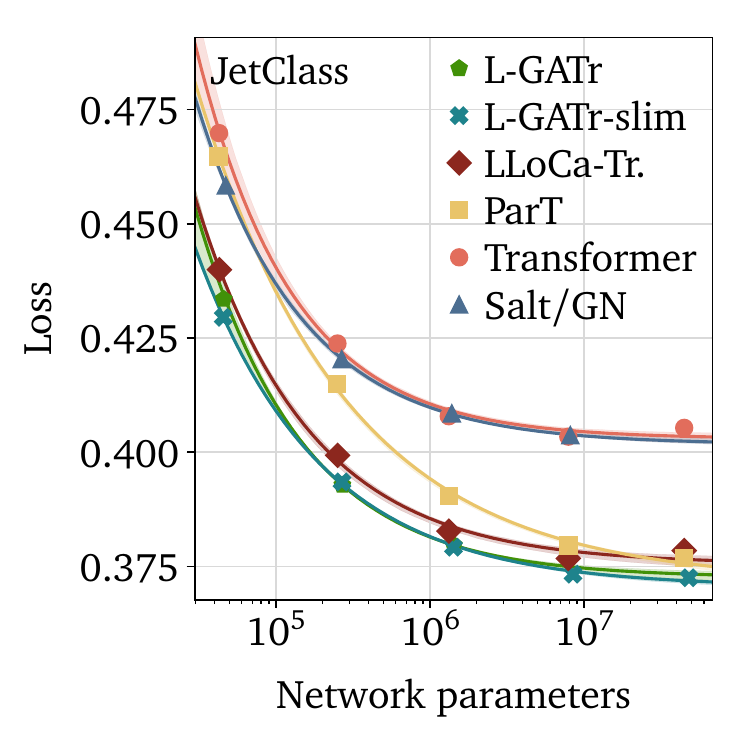}
    \includegraphics[width=0.485\linewidth,page=1]{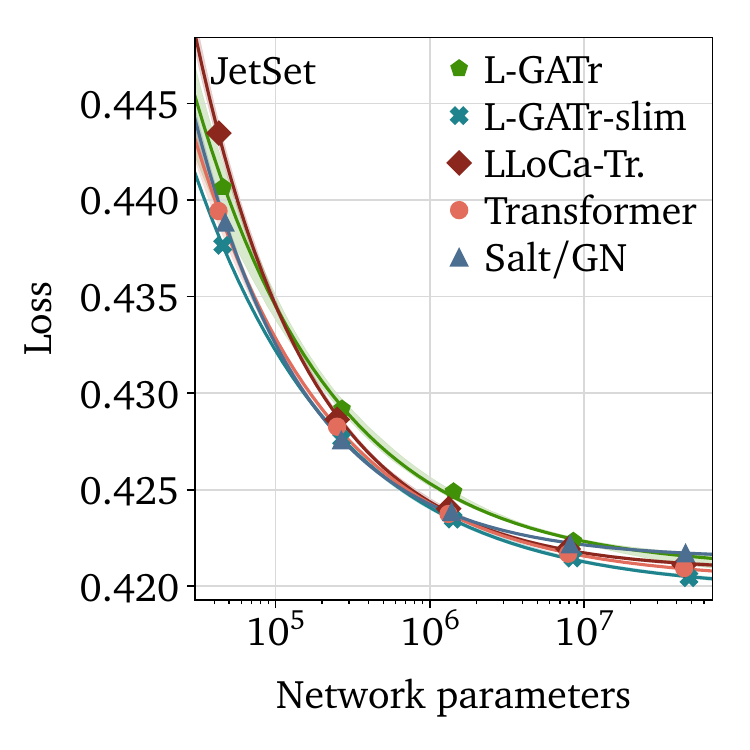}
    \caption{JetSet and JetClass performance scaling curves for the loss function for all the taggers, including the Salt/GN network. }
    \label{fig:perf_wsalt}
\end{figure}

Figure~\ref{fig:perf_wsalt} shows the test loss over the number of network parameters for the JetClass and JetSet datasets, this time including the Salt/GN transformer. Consistently with our expectations, the Salt/GN transformer is a variant of a standard transformer and shows similar scaling fits. We include the fit parameters and compare them with the vanilla transformer in Table~\ref{tab:scaling_exponents_wsalt}. We observe a slightly different scaling exponent on JetClass, while still reaching a similar $L_\infty$ test loss. The scaling on Jetset is the same within uncertainties, while we note a slightly better $L_\infty$ test loss for our transformer.

\begin{table}[t]
\centering
\begin{small} \begin{tabular}{lcccc}
  \toprule
  & \multicolumn{2}{c}{Exponent $\beta$} & \multicolumn{2}{c}{Asymptote $L_\infty$} \\
  & JetClass & JetSet & JetClass & JetSet \\
  \midrule
  Transformer &  $0.70_{-0.07}^{+0.07}$ & $0.50_{-0.02}^{+0.02}$ & $0.4032_{-0.0015}^{+0.0009}$ & $0.4207_{-0.0001}^{+0.0001}$ \\
  Salt/GN & $0.63_{-0.03}^{+0.04}$ & $0.60_{-0.03}^{+0.02}$ &  $0.4015_{-0.0004}^{+0.0009}$ & $0.4214_{-0.0002}^{+0.0001}$ \\
  \bottomrule
\end{tabular}
\end{small}
\caption{Exponent $\beta$ and asymptotic test loss $L_\infty$ of the scaling functions in Eq.\eqref{eq:scaling} of the transformer and the Salt/GN networks. We report the median and 10\%/90\% quantile uncertainties from our bootstrapping approach.}
\label{tab:scaling_exponents_wsalt}
\end{table}
\bibliographystyle{tepml}
\bibliography{tilman,literature}

@article{salt2025,
  author = {Jackson Barr and Diptaparna Biswas and Maxence Draguet and Philipp Gadow and Emil Haines and Osama Karkout and Dmitrii Kobylianskii and Wei Sheng Lai and Matthew Leigh and Nicholas Luongo and Ivan Oleksiyuk and Nikita Pond and Sébastien Rettie and Andrius Vaitkus and Samuel Van Stroud and Johannes Wagner},
  title = {Salt: Multimodal Multitask Machine Learning for High Energy Physics},
  journal = {Journal of Open Source Software},
  volume = {10},
  issue = {112},
  year = {2025},
  doi = {10.21105/joss.07217},
  url = {https://joss.theoj.org/papers/10.21105/joss.07217},
  issn = {2475-9066}
}

@article{Alioli:2010xd,
    author = "Alioli, Simone and Nason, Paolo and Oleari, Carlo and Re, Emanuele",
    title = "{A general framework for implementing NLO calculations in shower Monte Carlo programs: the POWHEG BOX}",
    eprint = "1002.2581",
    archivePrefix = "arXiv",
    primaryClass = "hep-ph",
    reportNumber = "DESY-10-018, SFB-CPP-10-22, IPPP-10-11, DCPT-10-22",
    doi = "10.1007/JHEP06(2010)043",
    journal = "JHEP",
    volume = "06",
    pages = "043",
    year = "2010"
}

@article{NNPDF:2014otw,
    author = "Ball, Richard D. and others",
    collaboration = "NNPDF",
    title = "{Parton distributions for the LHC Run II}",
    eprint = "1410.8849",
    archivePrefix = "arXiv",
    primaryClass = "hep-ph",
    reportNumber = "EDINBURGH-2014-15, IFUM-1034-FT, CERN-PH-TH-2013-253, OUTP-14-11P, CAVENDISH-HEP-14-11",
    doi = "10.1007/JHEP04(2015)040",
    journal = "JHEP",
    volume = "04",
    pages = "040",
    year = "2015"
}

@article{ATLAS:2025dkv,
    author = "Aad, Georges and others",
    collaboration = "ATLAS",
    title = "{Transforming jet flavour tagging at ATLAS}",
    eprint = "2505.19689",
    archivePrefix = "arXiv",
    primaryClass = "hep-ex",
    reportNumber = "CERN-EP-2025-103",
    doi = "10.1038/s41467-025-65059-6",
    journal = "Nature Commun.",
    volume = "17",
    number = "1",
    pages = "541",
    year = "2026"
}

@article{Louppe:2017ipp,
    author = "Louppe, Gilles and Cho, Kyunghyun and Becot, Cyril and Cranmer, Kyle",
    title = "{QCD-Aware Recursive Neural Networks for Jet Physics}",
    eprint = "1702.00748",
    archivePrefix = "arXiv",
    primaryClass = "hep-ph",
    doi = "10.1007/JHEP01(2019)057",
    journal = "JHEP",
    volume = "01",
    pages = "057",
    year = "2019"
}

@article{Cacciari:2011ma,
    author = "Cacciari, Matteo and Salam, Gavin P. and Soyez, Gregory",
    title = "{FastJet User Manual}",
    eprint = "1111.6097",
    archivePrefix = "arXiv",
    primaryClass = "hep-ph",
    reportNumber = "CERN-PH-TH-2011-297",
    doi = "10.1140/epjc/s10052-012-1896-2",
    journal = "Eur. Phys. J. C",
    volume = "72",
    pages = "1896",
    year = "2012"
}

@article{Alwall:2014hca,
    author = "Alwall, J. and Frederix, R. and Frixione, S. and Hirschi, V. and Maltoni, F. and Mattelaer, O. and Shao, H. -S. and Stelzer, T. and Torrielli, P. and Zaro, M.",
    title = "{The automated computation of tree-level and next-to-leading order differential cross sections, and their matching to parton shower simulations}",
    eprint = "1405.0301",
    archivePrefix = "arXiv",
    primaryClass = "hep-ph",
    reportNumber = "CERN-PH-TH-2014-064, CP3-14-18, LPN14-066, MCNET-14-09, ZU-TH-14-14",
    doi = "10.1007/JHEP07(2014)079",
    journal = "JHEP",
    volume = "07",
    pages = "079",
    year = "2014"
}

@InProceedings{dehaan2023euclidean,
  title = 	 {Euclidean, Projective, Conformal: Choosing a Geometric Algebra for Equivariant Transformers},
  author =       {de Haan, Pim and Cohen, Taco and Brehmer, Johann},
  booktitle = 	 {Proceedings of The 27th International Conference on Artificial Intelligence and Statistics},
  pages = 	 {3088--3096},
  year = 	 {2024},
  volume = 	 {238},
  series = 	 {Proceedings of Machine Learning Research},
  month = 	 {02--04 May},
  publisher =    {PMLR},
      eprint={2311.04744},
      archivePrefix={arXiv},
      primaryClass={cs.LG}
}

@article{Patel:2026zbq,
    author = "Patel, Pahal D. and Ganguly, Sanmay",
    title = "{Explainable AI for Jet Tagging: A Comparative Study of GNNExplainer, GNNShap, and GradCAM for Jet Tagging in the Lund Jet Plane}",
    eprint = "2604.25885",
    archivePrefix = "arXiv",
    primaryClass = "hep-ph",
    month = "4",
    year = "2026"
}

@article{Flek:2026vof,
    author = "Flek, Lucie and Jungs, Philipp Alexander and Karimi, Akbar and Saala, Timo and Schmid, Alexander and Schott, Matthias and Soldin, Philipp and Wiebusch, Christopher and Willemsen, Ulrich",
    title = "{Uncovering Hidden Systematics in Neural Network Models for High Energy Physics}",
    eprint = "2605.07470",
    archivePrefix = "arXiv",
    primaryClass = "cs.LG",
    month = "5",
    year = "2026"
}

@article{Agarwal:2026uqw,
    author = "Agarwal, Jay and Khare, Siddharth and Kumar, Dhruv",
    title = "{What Do Lorentz-Equivariant Jet Taggers Learn?}",
    eprint = "2606.21790",
    archivePrefix = "arXiv",
    primaryClass = "cs.LG",
    month = "6",
    year = "2026"
}

@article{Vigl:2026ppx,
    author = "Vigl, Matthias and Hartman, Nicole and Kagan, Michael and Heinrich, Lukas",
    title = "{Neural Scaling Laws for Boosted Jet Tagging}",
    eprint = "2602.15781",
    archivePrefix = "arXiv",
    primaryClass = "hep-ex",
    month = "2",
    year = "2026"
}

@article{ATLAS:2024rua,
    author = "Aad, Georges and others",
    collaboration = "ATLAS",
    title = "{Accuracy versus precision in boosted top tagging with the ATLAS detector}",
    eprint = "2407.20127",
    archivePrefix = "arXiv",
    primaryClass = "hep-ex",
    reportNumber = "CERN-EP-2024-159",
    doi = "10.1088/1748-0221/19/08/P08018",
    journal = "JINST",
    volume = "19",
    number = "08",
    pages = "P08018",
    year = "2024"
}

@inproceedings{ruhe2023clifford,
  title={Clifford Group Equivariant Neural Networks},
  author={Ruhe, David and Brandstetter, Johannes and Forr\'e, Patrick},
  booktitle = {{Advances in Neural Information Processing Systems}},
  volume={37},
  year={2023},
    eprint = "2305.11141",
    archivePrefix = "arXiv",
    primaryClass = "cs.LG",
    month = "5"
}

@article{Qu:2019gqs,
    author = "Qu, Huilin and Gouskos, Loukas",
    title = "{ParticleNet: Jet Tagging via Particle Clouds}",
    eprint = "1902.08570",
    archivePrefix = "arXiv",
    primaryClass = "hep-ph",
    doi = "10.1103/PhysRevD.101.056019",
    journal = "Phys. Rev. D",
    volume = "101",
    number = "5",
    pages = "056019",
    year = "2020"
}

@article{Qu:2022mxj,
    author = "Qu, Huilin and Li, Congqiao and Qian, Sitian",
    title = "{Particle Transformer for Jet Tagging}",
    eprint = "2202.03772",
    archivePrefix = "arXiv",
    primaryClass = "hep-ph",
    month = "2",
    year = "2022"
}

@article{Gong:2022lye,
    author = "Gong, Shiqi and Meng, Qi and Zhang, Jue and Qu, Huilin and Li, Congqiao and Qian, Sitian and Du, Weitao and Ma, Zhi-Ming and Liu, Tie-Yan",
    title = "{An efficient Lorentz equivariant graph neural network for jet tagging}",
    eprint = "2201.08187",
    archivePrefix = "arXiv",
    primaryClass = "hep-ph",
    doi = "10.1007/JHEP07(2022)030",
    journal = "JHEP",
    volume = "07",
    pages = "030",
    year = "2022"
}

@article{Bogatskiy:2023nnw,
    author = "Bogatskiy, Alexander and Hoffman, Timothy and Miller, David W. and Offermann, Jan T. and Liu, Xiaoyang",
    title = "{Explainable equivariant neural networks for particle physics: PELICAN}",
    eprint = "2307.16506",
    archivePrefix = "arXiv",
    primaryClass = "hep-ph",
    doi = "10.1007/JHEP03(2024)113",
    journal = "JHEP",
    volume = "03",
    pages = "113",
    year = "2024"
}

@article{Sjostrand:2014zea,
    author = {Sj{\"o}strand, Torbj{\"o}rn and Ask, Stefan and Christiansen, Jesper R. and Corke, Richard and Desai, Nishita and Ilten, Philip and Mrenna, Stephen and Prestel, Stefan and Rasmussen, Christine O. and Skands, Peter Z.},
    title = "{An introduction to PYTHIA 8.2}",
    eprint = "1410.3012",
    archivePrefix = "arXiv",
    primaryClass = "hep-ph",
    reportNumber = "LU-TP-14-36, MCNET-14-22, CERN-PH-TH-2014-190, FERMILAB-PUB-14-316-CD, DESY-14-178, SLAC-PUB-16122",
    doi = "10.1016/j.cpc.2015.01.024",
    journal = "Comput. Phys. Commun.",
    volume = "191",
    pages = "159--177",
    year = "2015"
}

@article{deFavereau:2013fsa,
    author = "de Favereau, J. and Delaere, C. and Demin, P. and Giammanco, A. and Lema{\^\i}tre, V. and Mertens, A. and Selvaggi, M.",
    collaboration = "DELPHES 3",
    title = "{DELPHES 3, A modular framework for fast simulation of a generic collider experiment}",
    eprint = "1307.6346",
    archivePrefix = "arXiv",
    primaryClass = "hep-ex",
    doi = "10.1007/JHEP02(2014)057",
    journal = "JHEP",
    volume = "02",
    pages = "057",
    year = "2014"
}

@article{Wu:2024thh,
    author = "Wu, Yifan and Wang, Kun and Li, Congqiao and Qu, Huilin and Zhu, Jingya",
    title = "{Jet tagging with more-interaction particle transformer*}",
    eprint = "2407.08682",
    archivePrefix = "arXiv",
    primaryClass = "hep-ph",
    doi = "10.1088/1674-1137/ad7f3d",
    journal = "Chin. Phys. C",
    volume = "49",
    number = "1",
    pages = "013110",
    year = "2025"
}

@article{Krause:2025qnl,
    author = "Krause, Claudius and Wang, Daohan and Winterhalder, Ramon",
    title = "{BitHEP {\textemdash} The limits of low-precision ML in HEP}",
    eprint = "2504.03387",
    archivePrefix = "arXiv",
    primaryClass = "hep-ph",
    reportNumber = "HEPHY-ML-25-02",
    doi = "10.21468/SciPostPhys.20.2.038",
    journal = "SciPost Phys.",
    volume = "20",
    number = "2",
    pages = "038",
    year = "2026"
}

@article{Rai:2025cog,
    author = "Rai, Saurabh and Prisha and Kumar, Jitendra",
    title = "{Investigating 1-bit quantization in transformer-based top tagging}",
    eprint = "2508.07431",
    archivePrefix = "arXiv",
    primaryClass = "hep-ph",
    doi = "10.1140/epjc/s10052-026-15985-6",
    journal = "Eur. Phys. J. C",
    volume = "86",
    number = "7",
    pages = "758",
    year = "2026"
}

@article{Esmail:2025kii,
    author = "Esmail, Waleed and Hammad, Ahmed and Nojiri, Mihoko",
    title = "{IAFormer: Interaction-Aware Transformer network for collider data analysis}",
    eprint = "2505.03258",
    archivePrefix = "arXiv",
    primaryClass = "hep-ph",
    doi = "10.21468/SciPostPhys.20.4.108",
    journal = "SciPost Phys.",
    volume = "20",
    number = "4",
    pages = "108",
    year = "2026"
}

@article{Komiske:2018cqr,
    author = "Komiske, Patrick T. and Metodiev, Eric M. and Thaler, Jesse",
    title = "{Energy Flow Networks: Deep Sets for Particle Jets}",
    eprint = "1810.05165",
    archivePrefix = "arXiv",
    primaryClass = "hep-ph",
    reportNumber = "MIT-CTP 5064",
    doi = "10.1007/JHEP01(2019)121",
    journal = "JHEP",
    volume = "01",
    pages = "121",
    year = "2019"
}

@article{Bogatskiy:2022czk,
    author = "Bogatskiy, Alexander and Hoffman, Timothy and Miller, David W. and Offermann, Jan T.",
    title = "{PELICAN: Permutation Equivariant and Lorentz Invariant or Covariant Aggregator Network for Particle Physics}",
    eprint = "2211.00454",
    archivePrefix = "arXiv",
    primaryClass = "hep-ph",
    month = "11",
    year = "2022"
}

@article{Bhimji:2025isp,
    author = "Bhimji, Wahid and Harris, Chris and Mikuni, Vinicius and Nachman, Benjamin",
    title = "{Foundation model framework for all tasks involving jet physics}",
    eprint = "2510.24066",
    archivePrefix = "arXiv",
    primaryClass = "hep-ph",
    doi = "10.1103/knmd-f5jm",
    journal = "Phys. Rev. D",
    volume = "113",
    number = "3",
    pages = "032020",
    year = "2026"
}

@article{Nachman:2022emq,
	author       = {Nachman, Ben and others},
	title        = {{Jets and Jet Substructure at Future Colliders}},
	year         = 2022,
	journal      = {Front. in Phys.},
	volume       = 10,
	pages        = 897719,
	doi          = {10.3389/fphy.2022.897719},
	eprint       = {2203.07462},
	archiveprefix = {arXiv},
	primaryclass = {hep-ph},
	reportnumber = {FERMILAB-PUB-22-186-SCD-T}
}

@article{Cogan:2014oua,
    author = "Cogan, Josh and Kagan, Michael and Strauss, Emanuel and Schwarztman, Ariel",
    title = "{Jet-Images: Computer Vision Inspired Techniques for Jet Tagging}",
    eprint = "1407.5675",
    archivePrefix = "arXiv",
    primaryClass = "hep-ph",
    doi = "10.1007/JHEP02(2015)118",
    journal = "JHEP",
    volume = "02",
    pages = "118",
    year = "2015"
}

@article{Baldi:2014kfa,
    author = "Baldi, Pierre and Sadowski, Peter and Whiteson, Daniel",
    title = "{Searching for Exotic Particles in High-Energy Physics with Deep Learning}",
    eprint = "1402.4735",
    archivePrefix = "arXiv",
    primaryClass = "hep-ph",
    doi = "10.1038/ncomms5308",
    journal = "Nature Commun.",
    volume = "5",
    pages = "4308",
    year = "2014"
}

@article{deOliveira:2015xxd,
    author = "de Oliveira, Luke and Kagan, Michael and Mackey, Lester and Nachman, Benjamin and Schwartzman, Ariel",
    title = "{Jet-images \textemdash{} deep learning edition}",
    eprint = "1511.05190",
    archivePrefix = "arXiv",
    primaryClass = "hep-ph",
    doi = "10.1007/JHEP07(2016)069",
    journal = "JHEP",
    volume = "07",
    pages = "069",
    year = "2016"
}

@article{Gallicchio:2010sw,
    author = "Gallicchio, Jason and Schwartz, Matthew D.",
    title = "{Seeing in Color: Jet Superstructure}",
    eprint = "1001.5027",
    archivePrefix = "arXiv",
    primaryClass = "hep-ph",
    doi = "10.1103/PhysRevLett.105.022001",
    journal = "Phys. Rev. Lett.",
    volume = "105",
    pages = "022001",
    year = "2010"
}

@article{Qiu:2022xvr,
    author = "Qiu, Shikai and Han, Shuo and Ju, Xiangyang and Nachman, Benjamin and Wang, Haichen",
    title = "{Holistic approach to predicting top quark kinematic properties with the covariant particle transformer}",
    eprint = "2203.05687",
    archivePrefix = "arXiv",
    primaryClass = "hep-ph",
    doi = "10.1103/PhysRevD.107.114029",
    journal = "Phys. Rev. D",
    volume = "107",
    number = "11",
    pages = "114029",
    year = "2023"
}

@article{Qiu:2023ihi,
    author = "Qiu, Shikai and Han, Shuo and Ju, Xiangyang and Nachman, Benjamin and Wang, Haichen",
    title = "{Parton labeling without matching: unveiling emergent labelling capabilities in regression models}",
    eprint = "2304.09208",
    archivePrefix = "arXiv",
    primaryClass = "hep-ph",
    doi = "10.1140/epjc/s10052-023-11809-z",
    journal = "Eur. Phys. J. C",
    volume = "83",
    number = "7",
    pages = "622",
    year = "2023"
}

@article{Birk:2025fbs,
    author = "Birk, Joschka and Hallin, Anna and Kasieczka, Gregor and Madzharova, Nikol and Pang, Ian and Shih, David",
    title = "{Enhancing next token prediction based pre-training for jet foundation models}",
    eprint = "2512.04149",
    archivePrefix = "arXiv",
    primaryClass = "hep-ph",
    month = "12",
    year = "2025"
}

@article{wang2023bitnet,
  title   = {BitNet: 1-bit Pre-training for Large Language Models},
  author={Wang, Hongyu and Ma, Shuming and Dong, Li and Huang, Shaohan and Wang, Huaijie and Ma, Lingxiao and Yang, Fan and Wang, Ruiping and Wu, Yi and Wei, Furu},
  journal = {Journal of Machine Learning Research},
    eprint = "2310.11453",
    archivePrefix = "arXiv",
    primaryClass = "cs.LG",
  year    = {2025},
  volume  = {26},
  number  = {125},
  pages   = {1--29},
  url     = {http://jmlr.org/papers/v26/24-2050.html}
}

@INPROCEEDINGS{6757323,
  author={Horowitz, Mark},
  booktitle={2014 IEEE International Solid-State Circuits Conference Digest of Technical Papers (ISSCC)}, 
  title={1.1 Computing's energy problem (and what we can do about it)}, 
  year={2014},
  volume={},
  number={},
  pages={10-14},
  doi={10.1109/ISSCC.2014.6757323}}

@inproceedings{openequivariance,
author={Vivek Bharadwaj and Austin Glover and Aydin Buluc and James Demmel},
title={An Efficient Sparse Kernel Generator for O(3)-Equivariant Deep Networks}, 
booktitle = {SIAM Conference on Applied and Computational Discrete Algorithms (ACDA25)},
url={https://arxiv.org/abs/2501.13986},
publisher={Society for Industrial and Applied Mathematics},
      eprint={2501.13986},
      archivePrefix={arXiv},
      primaryClass={cs.LG},
year={2025}
}

@misc{cuequivariance,
  author       = {{NVIDIA Corporation}},
  title        = {{cuEquivariance}: {CUDA}-accelerated equivariant operations},
  year         = {2024},
  version      = {0.9.0},
  howpublished = {\url{https://github.com/NVIDIA/cuEquivariance}},
  note         = {Documentation: \url{https://docs.nvidia.com/cuda/cuequivariance/}}
}

@inproceedings{Glorot:2010init,
    author        = "Glorot, Xavier and Bengio, Yoshua",
    title         = "{Understanding the difficulty of training deep feedforward neural networks}",
    booktitle     = "{Proceedings of the Thirteenth International Conference on Artificial Intelligence and Statistics (AISTATS)}",
    editor        = "Teh, Yee Whye and Titterington, Mike",
    series        = "Proceedings of Machine Learning Research",
    publisher     = "PMLR",
    volume        = "9",
    pages         = "249--256",
    year          = "2010"
}

@inproceedings{He:2015rectifiers,
    author        = "He, Kaiming and Zhang, Xiangyu and Ren, Shaoqing and Sun, Jian",
    title         = "{Delving Deep into Rectifiers: Surpassing Human-Level Performance on ImageNet Classification}",
    booktitle     = "{2015 IEEE International Conference on Computer Vision (ICCV)}",
    doi           = "10.1109/ICCV.2015.123",
    eprint        = "1502.01852",
    archivePrefix = "arXiv",
    primaryClass  = "cs.CV",
    pages         = "1026--1034",
    year          = "2015"
}

@techreport{CMS-DP-2024-066,
      collaboration = "CMS",
      author = "Gevorgyan, Arzunik and others",
    institution = "CMS",
      type          = "{CMS Detector Performance Note}",
      number        = "CMS-DP-2024-066",
      title         = "{A unified approach for jet tagging in Run 3 at
                       $\sqrt{s}$=13.6 TeV in CMS}",
      year          = "2024",
      url           = "https://cds.cern.ch/record/2904702",
}

@techreport{CMS-DP-2025-081,
      collaboration = "CMS",
      author = "Gevorgyan, Arzunik and others",
    institution = "CMS",
      type          = "{CMS Detector Performance Note}",
      number        = "CMS-DP-2025-081",
      title         = "{Flavour tagging performance of the updated Unified
                       Particle Transformer algorithm with the CMS experiment at
                       $\sqrt{s}$=13.6 TeV}",
      year          = "2025",
      url           = "https://cds.cern.ch/record/2948917",
}

@techreport{CMS-DP-2026-104,
      collaboration = "CMS",
      author = "Gevorgyan, Arzunik and others",
    institution = "CMS",
      type          = "{CMS Detector Performance Note}",
      number        = "CMS-DP-2026-104",
      title         = "{The Global Particle Transformer for boosted-jet tagging
                       and mass regression: Development and performance from v1 to
                       v3}",
      year          = "2026",
      url           = "https://cds.cern.ch/record/2966239",
}

@techreport{ATL-PHYS-PUB-2026-001,
      collaboration = "ATLAS",
    author = "Aad, Georges and others",
    institution = "ATLAS",
      type          = "{ATLAS Public Note}",
      title         = "{GN3: Multi-task, Multi-modal Transformers for Jet Flavour
                       Tagging in ATLAS}",
      number        = "ATL-PHYS-PUB-2026-001",
      year          = "2026",
      url           = "https://cds.cern.ch/record/2953652",
}

@techreport{ATL-PHYS-PUB-2026-013,
      collaboration = "ATLAS",
    author = "Aad, Georges and others",
    institution = "ATLAS",
      type          = "{ATLAS Public Note}",
      title         = "{GN3X: Improved Transformer-based Tagger for Boosted Higgs
                       Bosons in ATLAS}",
      number        = "ATL-PHYS-PUB-2026-013",
      year          = "2026",
      url           = "https://cds.cern.ch/record/2965597",
}

@article{ATLAS:2026vyw,
    collaboration = "ATLAS",
    author = "Aad, Georges and others",
    institution = "ATLAS",
    title = "{GN3: Multi-task, Multi-modal Transformers for Jet Flavour Tagging in ATLAS}",
    reportNumber = "ATL-PHYS-PUB-2026-001",
    year = "2026"
}

@article{Barr:2025djz,
    author = "Barr, Jackson and others",
    title = "{Salt: Multimodal Multitask Machine Learning for High Energy Physics}",
    doi = "10.21105/joss.07217",
    journal = "J. Open Source Softw.",
    volume = "10",
    number = "112",
    pages = "7217",
    year = "2025"
}

@article{Breso-Pla:2026tlz,
    author = "Breso-Pla, Victor and Greif, Kevin and Mikuni, Vinicius and Nachman, Benjamin and Plehn, Tilman and Wamorkar, Tanvi and Whiteson, Daniel",
    title = "{Explicit or Implicit? Encoding Physics at the Precision Frontier}",
    eprint = "2603.08802",
    archivePrefix = "arXiv",
    primaryClass = "hep-ph",
    month = "3",
    year = "2026"
}

@article{Kuntz:2026kuv,
    author = {Kuntz, Rebecca Maria and Plehn, Tilman and Sch{\"a}fer, Bj{\"o}rn Malte and Schosser, Benedikt and Vent, Sophia},
    title = "{The Latent Information Geometry of Jet Classification}",
    eprint = "2603.02310",
    archivePrefix = "arXiv",
    primaryClass = "hep-ph",
    month = "3",
    year = "2026"
}

@article{Petitjean:2025zjf,
    author = {Petitjean, Antoine and Plehn, Tilman and Spinner, Jonas and K{\"o}the, Ullrich},
    title = "{Economical Jet Taggers -- Equivariant, Slim, and Quantized}",
    eprint = "2512.17011",
    archivePrefix = "arXiv",
    primaryClass = "hep-ph",
    reportNumber = "IPPP/25/93",
    month = "12",
    year = "2025"
}

@article{Villadamigo:2025our,
    author = "Villadamigo, Javier Mari{\~n}o and Frederix, Rikkert and Plehn, Tilman and Vitos, Timea and Winterhalder, Ramon",
    title = "{FASTColor -- Full-color Amplitude Surrogate Toolkit for QCD}",
    eprint = "2509.07068",
    archivePrefix = "arXiv",
    primaryClass = "hep-ph",
    reportNumber = "TIF-UNIMI-2025-18",
    doi = "10.21468/SciPostPhys.21.1.001",
    journal = "SciPost Phys.",
    volume = "21",
    pages = "001",
    year = "2026"
}

@article{Favaro:2025pgz,
    author = "Favaro, Luigi and Gerhartz, Gerrit and Hamprecht, Fred A. and Lippmann, Peter and Pitz, Sebastian and Plehn, Tilman and Qu, Huilin and Spinner, Jonas",
    title = "{Lorentz-Equivariance without Limitations}",
    eprint = "2508.14898",
    archivePrefix = "arXiv",
    primaryClass = "hep-ph",
    month = "8",
    year = "2025"
}

@article{Vent:2025ddm,
    author = "Vent, Sophia and Winterhalder, Ramon and Plehn, Tilman",
    title = "{The Physics Behind ML-based Quark-Gluon Taggers}",
    eprint = "2507.21214",
    archivePrefix = "arXiv",
    primaryClass = "hep-ph",
    reportNumber = "TIF-UNIMI-2025-16",
    doi = "10.21468/SciPostPhys.20.3.084",
    journal = "SciPost Phys.",
    volume = "20",
    pages = "084",
    year = "2026"
}

@article{Brehmer:2024yqw,
    author = "Brehmer, Johann and Bres{\'o}, V{\'\i}ctor and de Haan, Pim and Plehn, Tilman and Qu, Huilin and Spinner, Jonas and Thaler, Jesse",
    title = "{A Lorentz-equivariant transformer for all of the LHC}",
    eprint = "2411.00446",
    archivePrefix = "arXiv",
    primaryClass = "hep-ph",
    reportNumber = "MIT-CTP/5802",
    doi = "10.21468/SciPostPhys.19.4.108",
    journal = "SciPost Phys.",
    volume = "19",
    number = "4",
    pages = "108",
    year = "2025"
}

@article{Butter:2022xyj,
    author = "Butter, Anja and Dillon, Barry M. and Plehn, Tilman and Vogel, Lorenz",
    title = "{Performance versus resilience in modern quark-gluon tagging}",
    eprint = "2212.10493",
    archivePrefix = "arXiv",
    primaryClass = "hep-ph",
    doi = "10.21468/SciPostPhysCore.6.4.085",
    journal = "SciPost Phys. Core",
    volume = "6",
    pages = "085",
    year = "2023"
}

@article{Kasieczka:2019dbj,
    author = "Butter, Anja and others",
    editor = "Kasieczka, Gregor and Plehn, Tilman",
    title = "{The Machine Learning landscape of top taggers}",
    eprint = "1902.09914",
    archivePrefix = "arXiv",
    primaryClass = "hep-ph",
    doi = "10.21468/SciPostPhys.7.1.014",
    journal = "SciPost Phys.",
    volume = "7",
    pages = "014",
    year = "2019"
}

@article{Butter:2017cot,
    author = "Butter, Anja and Kasieczka, Gregor and Plehn, Tilman and Russell, Michael",
    title = "{Deep-learned Top Tagging with a Lorentz Layer}",
    eprint = "1707.08966",
    archivePrefix = "arXiv",
    primaryClass = "hep-ph",
    doi = "10.21468/SciPostPhys.5.3.028",
    journal = "SciPost Phys.",
    volume = "5",
    number = "3",
    pages = "028",
    year = "2018"
}

@article{Kasieczka:2017nvn,
    author = "Kasieczka, Gregor and Plehn, Tilman and Russell, Michael and Schell, Torben",
    title = "{Deep-learning Top Taggers or The End of QCD?}",
    eprint = "1701.08784",
    archivePrefix = "arXiv",
    primaryClass = "hep-ph",
    reportNumber = "MCNET-17-07",
    doi = "10.1007/JHEP05(2017)006",
    journal = "JHEP",
    volume = "05",
    pages = "006",
    year = "2017"
}

\end{document}